\pdfoutput=1
\documentclass[cernpreprint, atlasdraft=false, UKenglish, paper=a4, texmf, orcidlogo]{atlasdoc}
 
\usepackage[subcaption]{atlaspackage}
\usepackage{atlasbiblatex}
 
\usepackage{atlasphysics}

\graphicspath{{logos/}{figures/}}
 
\usepackage{ANA-EXOT-2019-28-PAPER-defs}

\AtlasTitle{Search for an axion-like particle with forward proton scattering in association with photon pairs at ATLAS}

\AtlasAbstract{
A search for forward proton scattering in association with light-by-light scattering mediated by an axion-like particle is presented, using the
ATLAS Forward Proton spectrometer to detect scattered protons and the central ATLAS detector to detect pairs of outgoing photons.
Proton--proton collision data recorded in 2017 at a centre-of-mass energy of $\sqrt{s} = 13$~TeV were analysed, corresponding to an integrated luminosity of 14.6~fb$^{-1}$.
A total of 441 candidate events were selected.
A search was made for a narrow resonance in the diphoton mass distribution, corresponding to an axion-like particle (ALP) with mass in the range $150$--1600~GeV.  No excess is observed above a smooth background.
Upper limits on the production cross section of a narrow resonance are set as a function of the mass,
and are interpreted as upper limits on the ALP production coupling constant, assuming 100\% decay branching
ratio into a photon pair. The inferred upper limit on the coupling constant is in the range  0.04--0.09~TeV$^{-1}$ at 95\% confidence level.
}

\AtlasRefCode{EXOT-2019-28}
 
\PreprintIdNumber{CERN-EP-2023-049}

\AtlasJournal{JHEP}
\AtlasJournalRef{JHEP 07 (2023) 234}
\AtlasDOI{10.1007/JHEP07(2023)234}

\hypersetup{pdftitle={ATLAS document},pdfauthor={The ATLAS Collaboration}}
 
\begin{document}
 
\maketitle

\section{Introduction}
 
The axion~\cite{PhysRevLett.40.223,PhysRevLett.40.279} is the name given to an as yet unobserved particle whose existence is proposed in order to explain the lack of CP violation in quantum chromodynamics~\cite{PhysRevLett.97.131801,PhysRevLett.124.081803,PhysRevLett.38.1440,PhysRevD.16.1791}. Such a particle must be very light and couple to two photons. Some extensions to the Standard Model (SM) predict the possibility of heavy particles which are axion-like in that they also couple to two photons. Such axion-like particles (ALPs) may be produced in two-photon collisions and subsequently decay into two photons~\cite{Knapen_2017,Baldenegro:2018hng,sc4_alpSDDD}; the ALP production process can therefore be identified as a resonant peak at the ALP mass value in the process \gggg, also known as light-by-light ($\gamma\gamma$) scattering.
 
Any high-energy source of real or weakly virtual photon pairs in collisions would in principle be capable of producing ALPs.
At the LHC, substantial fluxes of virtual photons can be generated by radiation from colliding high-energy protons or ions, and pairs of such photons can interact.
In processes of this kind, each incoming charged particle continues to travel close to its original direction.
The process \gggg\ can also occur through an intermediate fermion or $W$ boson box diagram.
It has been measured at the LHC in nucleus--nucleus collisions using lead-ion beams (Pb--Pb collisions)~\cite{HION-2016-05,CMS-FSQ-16-012,HION-2018-19,HION-2019-08}, where the \gggg\ cross section is enhanced because of the high nuclear charge.
These analyses also searched for an ALP mediated by $\gamma\gamma$ scattering with a mass $m_X$ up to 100~\GeV.
At higher diphoton masses, the effective $\gamma\gamma$ luminosity in $pp$ collisions surpasses that of Pb--Pb collisions~\cite{Bruce_2020}, although the scattering cross section is lower than at lower masses.
 
A pair of outgoing real photons, which is the signature for ALP production, can be detected in the central detector surrounding the interaction point, while suitable apparatus located in each of the forward directions may be used to detect the outgoing scattered protons (referred to as proton tagging) in $pp$ collisions~\cite{Baldenegro:2018hng}.
The production of lepton pairs by photon--photon interactions, $pp\to p(\gamma\gamma\to\ell^+\ell^-)p^{(*)}$, has been measured by ATLAS and CMS using forward-proton detectors~\cite{STDM-2018-16,CMS-PPS-17-001}.

\begin{figure}[hbpt]
\centering
\includegraphics[width=0.9\columnwidth]{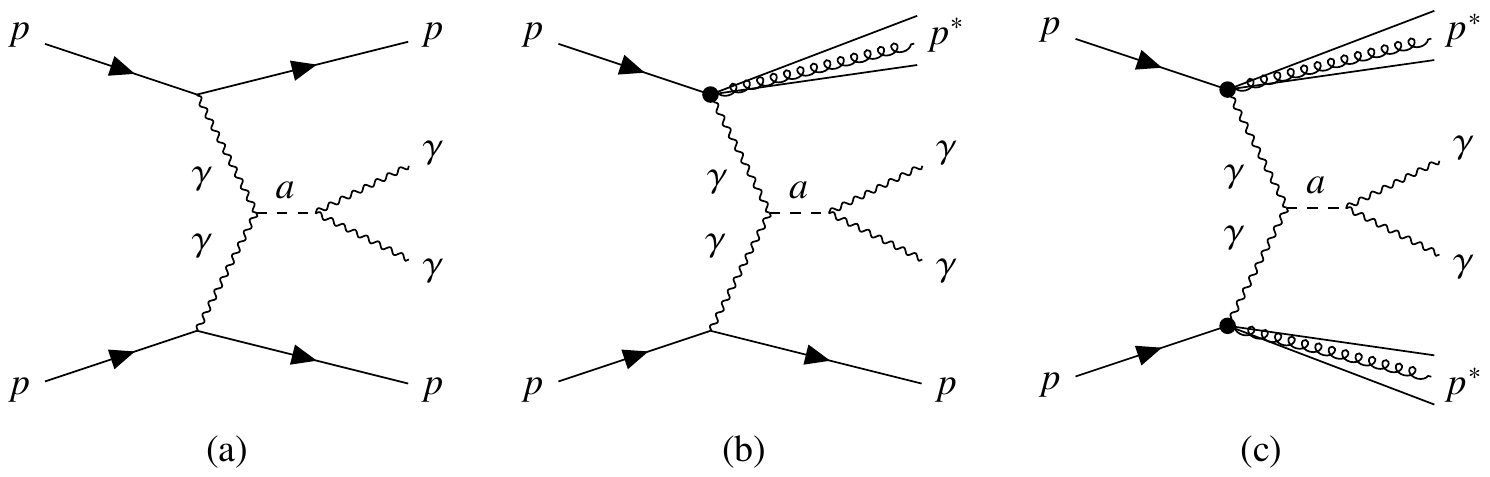}
\caption{Feynman diagrams for (a)~exclusive, (b)~single-dissociative, and (c)~double-dissociative light-by-light scattering with outgoing photon pairs mediated by an ALP denoted by $a$.}
\label{fig:feynman}
\end{figure}

This analysis presents a search for ALPs produced in light-by-light scattering in proton--proton ($pp$) collisions using the ATLAS detector. The target mass range is 150--1600~\GeV. Three possibilities for the reaction may be considered: the exclusive process \gggg\ measured as $pp\to pp\gamma\gamma$, and single- and double-dissociative processes (SD, DD) in which one or both protons ($p^*$) dissociate while radiating a virtual photon, as depicted in Figure~\ref{fig:feynman}.
While the undissociated proton may be tagged, the dissociated proton is in practice not measurable.
 
An exclusive signal search has been performed by the CMS--TOTEM collaborations in 9.4~fb$^{-1}$ of $pp$ collisions at $\sqrt{s}=13$~\TeV, making use of double proton tagging in a diphoton mass range of approximately 900--1800~\GeV~\cite{CMS-EXO-18-014}. A previous inclusive diphoton resonance search with ATLAS targeted the mass range 160--1160~GeV using a higher integrated luminosity, 139~fb$^{-1}$~\cite{HIGG-2018-27}.
The ATLAS search presented here uses 14.6~fb$^{-1}$ of 13~\TeV\ $pp$ collision data and requires at least one tagged proton,
giving a more specific measurement of the exclusive and SD processes with reduced background,
covering a mass range 150--1600~\GeV, and with higher experimental acceptance than in the pure double-tagging case.

\section{Experimental set-up}
 
\newcommand{\AtlasCoordFootnote}{
ATLAS uses a right-handed coordinate system with its origin at the nominal interaction point (IP)
in the centre of the detector and the \(z\)-axis along the beam pipe.
The region $z\geq 0$ is referred to as the A-side, and $z< 0$ is referred to as the C-side.
The \(x\)-axis points from the IP to the centre of the LHC ring,
and the \(y\)-axis points upwards.
Cylindrical coordinates \((r,\phi)\) are used in the transverse plane,
\(\phi\) being the azimuthal angle around the \(z\)-axis.
The pseudorapidity is defined in terms of the polar angle \(\theta\) as \(\eta = -\ln \tan(\theta/2)\).
Angular distance is measured in units of \(\Delta R \equiv \sqrt{(\Delta\eta)^{2} + (\Delta\phi)^{2}}\).}
 
The ATLAS experiment~\cite{PERF-2007-01} at the LHC is a multipurpose particle detector with
forward/backward-symmetric cylindrical geometry and near \(4\pi\) coverage in solid angle.\footnote{\AtlasCoordFootnote}
It consists of an inner tracking detector surrounded by a thin superconducting solenoid providing a \qty{2}{\tesla} axial magnetic field, electromagnetic and hadron calorimeters, and a muon spectrometer.
The inner tracking detector covers the pseudorapidity range \(|\eta| < 2.5\).
It consists of silicon pixel, silicon microstrip, and transition radiation tracking detectors.
Lead/liquid-argon (LAr) sampling calorimeters provide electromagnetic (EM) energy measurements with high granularity.
A steel/scintillator-tile hadron calorimeter covers the central pseudorapidity range (\(|\eta| < 1.7\)).
The endcap and forward regions are instrumented with LAr calorimeters for both the EM and hadronic energy measurements up to \(|\eta| = 4.9\).
The muon spectrometer (not used in the present analysis) surrounds the calorimeters and is based on three large superconducting air-core toroidal magnets with eight coils each.
The field integral of the toroids ranges between \num{2.0} and \qty{6.0}{\tesla\metre} across most of the detector.
The muon spectrometer  includes a system of precision tracking chambers and fast detectors for triggering.
 
Forward scattered protons are detected in Roman Pot systems known as the ATLAS Forward Proton (AFP) spectrometer system~\cite{ATLAS-TDR-24,ATL-PHYS-PUB-2017-012}.  These  detectors are positioned near the outgoing proton beam and can be moved in the $x$-direction close to the beam as required.
The AFP spectrometer consists of four tracking units located at $z = \pm 205$~m and $\pm 217$~m. They are denoted as \textsc{Near} and \textsc{Far} stations, respectively, with the
$+z\left(-z\right)$ direction denoted as the A(C)-side.
Each station houses a silicon tracker comprising four planes of edgeless silicon pixel sensors~\cite{Lange:2015xla,Lange:2016dro,GarciaSciveres:2011zz,Zivkovic_2012}.
The sensors have $336 \times 80$ pixels with area $50 \times 250~\mu$m$^2$.
Each sensor is tilted by $14^\circ$ relative to the $x$-direction to improve hit efficiency and $x$-position resolution, resulting in an overall spatial resolution of $\sigma_x = 6~\mu$m~\cite{Grinstein:2016hmq}.
Data taking with the AFP system commences once the detectors are at a position where the innermost silicon edge is within 2~mm of the beam centre during stable running conditions.
The AFP alignment calibration was performed using beam loss monitors~\cite{PhysRevSTAB.15.051002,Zamantzas2007}, beam position monitors~\cite{PhysRevAccelBeams.20.081002}, and dimuon production events, $\gamma\gamma\to\mu^+\mu^-$~\cite{STDM-2018-16}.
 
A two-level trigger system is used to select events.
The first-level trigger is implemented in hardware and makes use of a subset of the detector information to accept events at a rate below \qty{100}{\kHz}.
It is followed by a software-based trigger that reduces the average accepted event rate to \qty{1}{\kHz}, depending on the data-taking conditions.
No trigger with an AFP signal was used because each AFP station reconstructs proton tracks with a probability of around 70\% per bunch crossing.
 
An extensive software suite~\cite{ATL-SOFT-PUB-2021-001} is used in data simulation, in the reconstruction and analysis of real and simulated data, in detector operations, and in the trigger and data acquisition systems of the experiment.
 
\section{Data and simulated event samples}
\label{sec:samples}
 
The dataset was collected in 2017 using $pp$ collisions at a centre-of-mass energy $\sqrt{s}= 13~\TeV$. The integrated luminosity was 14.6~fb$^{-1}$, corresponding to the time the AFP system was in operation together with the central detector and after data-quality requirements.
The average number of interactions per bunch crossing was 36.
The data were recorded using a diphoton trigger that required two clusters of EM calorimeter cells with transverse energies $E_\mathrm{T}=E\sin (\theta)$ above $35~\GeV$ and $25~\GeV$, respectively~\cite{ATL-DAQ-PUB-2018-002,TRIG-2018-05}, after which standard data-quality requirements were applied~\cite{DAPR-2018-01}.
For the forward-proton measurement, it was required that every AFP station had at least three operational silicon planes, with correct operation of the AFP data acquisition system~\cite{Kocian:2017ygu}.
 
Simulated signal events were produced using the \textsc{SuperChic}\,4.02 Monte Carlo (MC) generator~\cite{sc2_ref,sc3_ref,sc4_ref} for the exclusive signals and \textsc{SuperChic}\,4.14~\cite{sc4_alpSDDD} for the SD and DD signals.
ALP masses in the range $m_X=150$--1600~\GeV\ were considered, and for each $m_X$ value a sample was generated with the ALP-to-diphoton coupling constant set to\footnote{$f^{-1}$ is taken from Ref.~\cite{Baldenegro:2018hng}, and the input coupling constant parameter of \textsc{SuperChic} is $g_a=4f^{-1}$~\cite{sc3_ref,sc4_alpSDDD}.} $f^{-1}=0.05~\TeV^{-1}$, where the natural width of the ALP is $\Gamma=m_X^3/4\pi f^2$.
Generator-level selections were applied, requiring photons to have transverse momentum $\pt>20$~\GeV\ and $\abseta<2.4$; the diphoton rapidity is required to be $|\yyy|<2.4$.
The acceptance $A_0$ for signal events to pass these fiducial region requirements is in the range 60\%--75\%, depending on $m_X$.
Typically, the SD production cross section is approximately three times the exclusive cross section, while the DD and exclusive cross sections are similar (the DD contribution drops to a negligible level after applying the selection cuts, especially for the narrow $\xi$ range).
The SD and DD events were interfaced with \Pythia[8.307]~\cite{10.21468/SciPostPhysCodeb.8} for hadronisation of dissociated-proton systems.
 
To model the central-detector response, the signal samples were processed with a fast simulation~\cite{SOFT-2010-01} which uses a \GEANT-based simulation of the inner tracker and a parameterisation of the calorimeter response~\cite{ATL-PHYS-PUB-2010-013}.  The reconstruction of scattered protons in the data combines information from the AFP tracker and knowledge of the LHC magnet lattice~\cite{STDM-2018-16,Staszewski_2009}, which is used to calculate the proton transport from the IP.  The response of the AFP spectrometer in the MC samples was modelled by a fast simulation in which Gaussian smearing, based on the AFP spatial resolution, is applied to the generated proton position in each AFP station.
The track is then reconstructed according to the simulated positions and subsequently used in the proton reconstruction.
Effects of multiple $pp$ interactions in the same and neighbouring bunch crossings (pile-up), as seen in the central detector, are included in the MC samples by overlaying simulated minimum-bias events generated with \Pythia[8.210]~\cite{Sjostrand:2014zea} using the A3 set of tuned parameters~\cite{ATL-PHYS-PUB-2016-017}. The distribution of the average number of interactions per bunch crossing in the MC simulation is reweighted to that observed in the data. Before applying the photon identification criteria to simulated events, the shower shape parameters were corrected for small differences in their average values between data and simulation.
 
\section{Object and event selection}
 
Photon candidates were reconstructed from topological clusters~\cite{PERF-2014-07} of energy deposits in the ATLAS EM calorimeter and were calibrated as described in Ref.~\cite{EGAM-2018-01}.
To reduce the background from jets, photon candidates were required to fulfil \textit{tight} identification criteria based on shower shapes in the EM calorimeter and energy leakage into the hadronic calorimeter~\cite{PERF-2017-01, EGAM-2018-01}; these gave a typical identification efficiency above 90\% and misidentification efficiency of 20--40\%.
The photon isolation requirement for further rejection of jets used only the calorimeter information, because
track-based isolation would remove photon-induced signal events accompanied by tracks from other interactions in the same bunch crossing if their primary vertices were misidentified as the signal vertex.
The calorimeter isolation transverse energy $E_{\mathrm{T}}^{\text{iso }}$ was required to be smaller than $0.022 E_{\mathrm{T}}+2.45$~\GeV, where $E_{\mathrm{T}}^{\text{iso }}$ is defined as the sum of transverse energies of the positive-energy topological clusters within a cone of size $\Delta R=0.4$ around the photon candidate; the main part (``core'') of the photon shower was excluded, and a correction was applied for leakage of the photon shower from the core into the rest of the isolation cone. The contributions from the underlying event and pile-up were subtracted using the techniques described in Refs.~\cite{Cacciari_2008,Cacciari_2010}.
The event selection required at least two photon candidates with $\pt > 40~\GeV$ and $|\eta|<2.37$, matched to the online photon objects that triggered the event, excluding the barrel-to-endcap transition regions of the calorimeter, $1.37<|\eta|<1.52$.
The azimuthal misalignment between the pair of photons was required to be small, as determined by an acoplanarity $\acoyy = 1-|\Delta \phi_{\gamma\gamma}| /\pi < 0.01$.
 
Protons transported to the AFP by the beamline magnets leave hits in its silicon trackers, which are processed by per-plane clustering and per-station track-finding algorithms~\cite{Lange:2016dro}.
Tracks are reconstructed from clusters found in at least two planes in each station.
From the track spatial coordinates, the proton energy and momentum at the IP are inferred, using the known beam optics.
The result of the reconstruction is the fractional energy loss of the scattered proton, defined as $\xi = 1-E_\text{scattered} / E_\text{beam}$, where $E_\text{scattered}$ ($E_\text{beam}$) is the scattered (beam) proton energy.
The determination of $\xi$ from the AFP stations ($\xiAFP$) requires tracks in both the \textsc{Near} and \textsc{Far} stations.
A $\xiAFP$ calibration was performed in the earlier measurement of dimuon production events recorded during the same data taking~\cite{STDM-2018-16}.  The same calibration was used in the present analysis, and made use of
a matching between the central-detector measurement and the AFP measurement, as described below.
Events with at least one A-side or C-side tagged proton were accepted, for which
the $\xiAFP$ value was required to be within the range $[0.035, 0.08]$.
 
The $\xi$ value of the forward scattered proton can also be calculated independently of the forward protons, using the kinematics of the central photon pair, and is denoted by $\xiyy$.
It is determined from the diphoton mass $\myy$ and rapidity $\yyy$ by momentum conservation as $\xiyy^{\pm} = ( \myy / \sqrt{s} )\;\!\mathrm{e}^{ \pm \yyy }$, where $+$ ($-$) corresponds to the proton on the A(C)-side and negligible transverse momentum transfer to the protons is assumed.
If there was more than one proton candidate on the same AFP side (\textasciitilde10\% of the cases), the proton with $\xiAFP$ closest to $\xiyy$ was chosen, giving more than 99\% efficiency.
Proton-tagged diphoton candidates were then selected by requiring kinematic matching on at least one AFP side: $\left | \Delta\xi \right|=\left | \xiAFP - \xiyy \right| < \xi_\mathrm{th}$, where $\xi_\mathrm{th}=0.004+0.1\xiyy$.
The second term in $\xi_\mathrm{th}$ takes account of the uncertainty in the proton propagation through the beam optics between the central detector and the AFP spectrometer, which is proportional to $\xiyy$.
For $\myy<500$~\GeV\ ($\geq500$~\GeV), the exclusive and SD processes respectively account for about 60\% (40\%) and 40\% (60\%) of the selected signal events.
 
\section{Signal and background modelling}
 
\begin{figure}[tbp]
\centering
\begin{tabular}{cc}
\begin{minipage}[t]{0.47\columnwidth}
\centering
\includegraphics[clip, width=\textwidth]{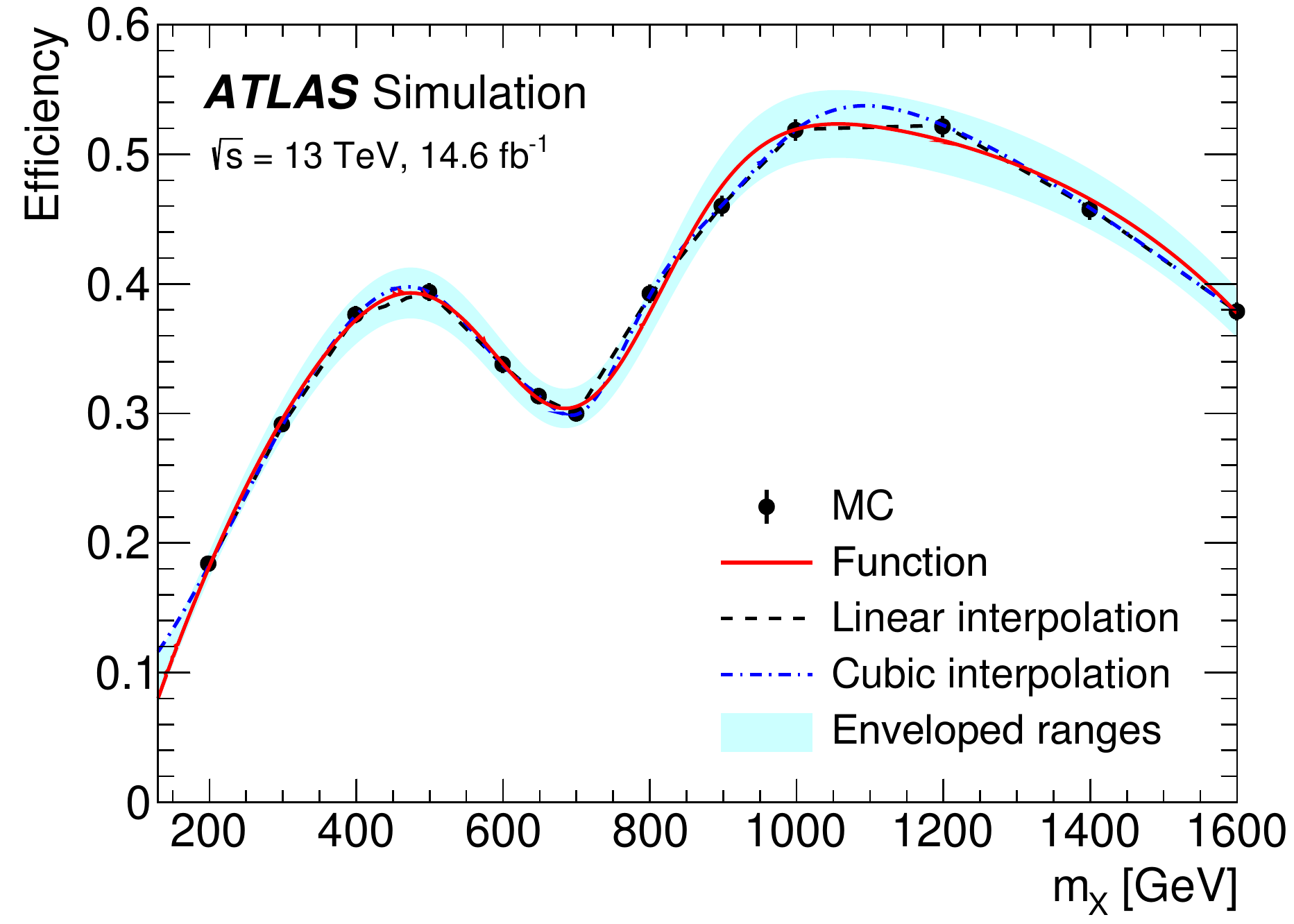}\\
(a)
\end{minipage} &
\begin{minipage}[t]{0.47\columnwidth}
\centering
\includegraphics[clip, width=\textwidth]{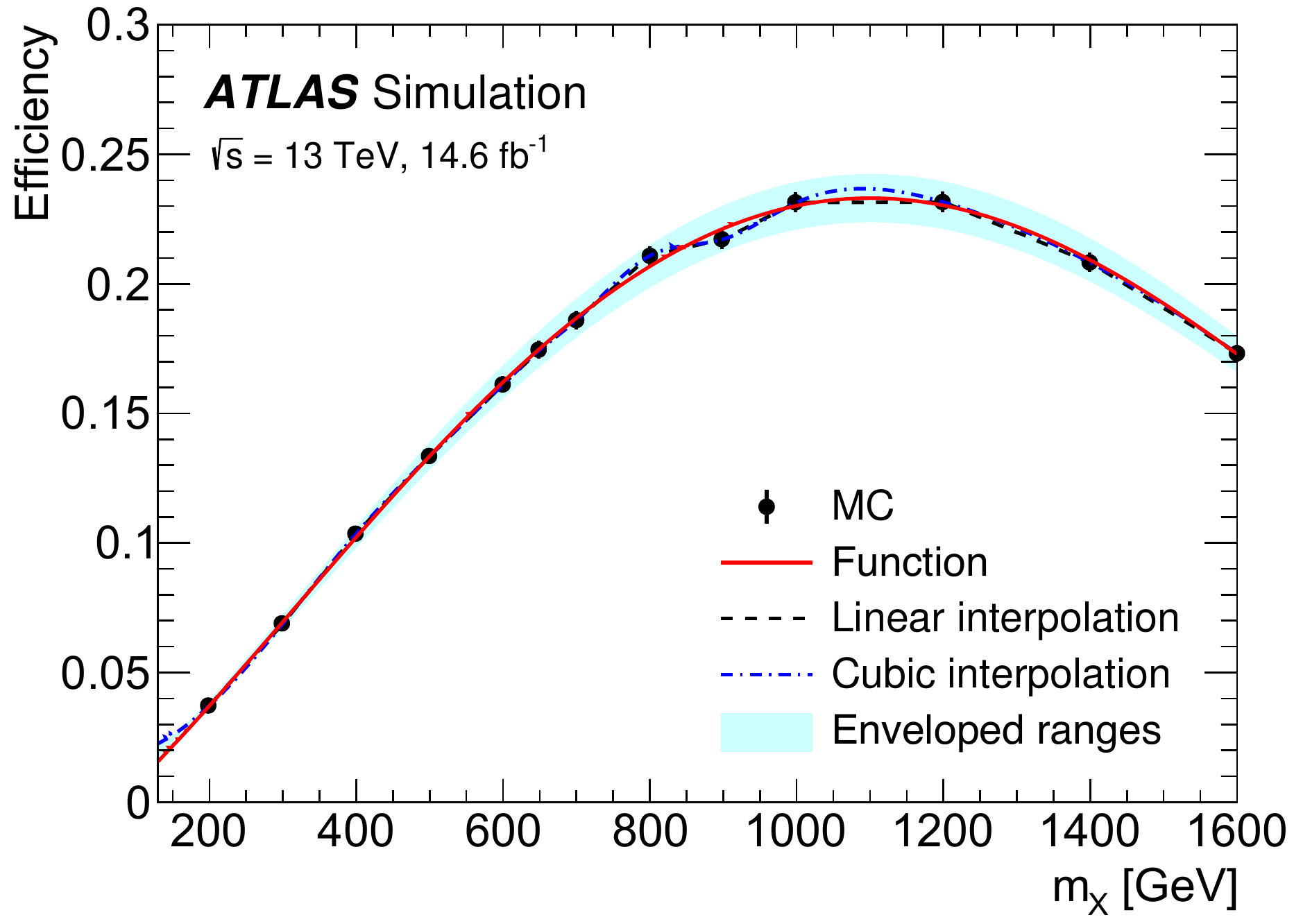}\\
(b)
\end{minipage}
\end{tabular}
\caption{\label{fig:signal_efficiency}
Signal selection efficiency as a function of ALP mass $m_X$ for the (a)~exclusive and (b)~single-dissociative process. The ratio of the number of selected events to the number of generated MC events is given (black points) and is parameterised by an analytic function (red solid line). The linear (black dashed line) and cubic (blue chain line) interpolations of the black points are used to derive the envelopes (cyan filled region) which are regarded as systematic uncertainties.
}
\end{figure}

For signal events, the diphoton invariant-mass distribution is expected to peak close to the mass of the ALP, with a spread given by the convolution of its intrinsic decay width and the experimental resolution.
For the ALP parameters of interest in this analysis, the ALP intrinsic decay width is narrow enough to be ignored, as inferred from the $m_X$ and $f^{-1}$ values input to the signal MC samples.
The experimental resolution of the invariant mass is modelled with a double-sided Crystal Ball (DSCB) function~\cite{HIGG-2014-04,Oreglia:1980cs}.
The six parameters of the DSCB function are determined using the simulated signal MC samples. Each parameter is expressed as a linear function of $m_X$. The uncertainty due to the fitting procedure for the linear function
is used as a systematic uncertainty for each parameter of the DSCB function.
The signal fiducial cross section, $\sigma_\mathrm{fid}$, and the signal selection efficiency are also modelled as functions of $m_X$ to derive exclusion upper limits for the cross section. The models are illustrated in Figure~\ref{fig:signal_efficiency}. In the exclusive process, the
combination of the AFP geometrical acceptance and the $\yyy$ distribution of the expected signal gives rise to a double-peaked mass structure in the exclusive signal efficiency, which is a consequence of the acceptance values integrated over the mass contours overlapping with the shaded region in Figure~\ref{fig:xiA_xiC_data}.\footnote{In Figure~\ref{fig:xiA_xiC_data},
the contours at 400 and 1000-1200~\GeV\ pass maximally through the shaded regions of acceptance, corresponding to peaks in the efficiency distribution, while the contours at 600-800~\GeV\ pass through a smaller acceptance region of acceptance, which generates a dip.}
 
The dominant source of background after the full event selection arises when a pair of photons (or hadronic jets misidentified as photons) is produced in a $pp$ interaction other than that giving the detected forward proton but within the same bunch crossing.
The protons originate from soft-scale events, in most cases single-diffractive processes.
These recorded events are collectively referred to as combinatorial backgrounds and are modelled using a fully data-driven method which follows the background estimation procedure.
An event sample, referred to as a “mixed-data” sample, was constructed by replacing the AFP data from a given event by that from one or more other data events, before the event selection. To maximise the statistical precision, this reassignment was performed using the method described in Ref.~\cite{STDM-2018-16}.
It uses each possible combination of diphoton and AFP data from different events within intervals of measured instantaneous luminosity, and suppresses any contribution from signal events and from background events with a single vertex.
The normal event selection was then applied to the mixed-data sample, and the events were subsequently inversely weighted by the number of reassignments that were taken for a given diphoton event.
This maximal use of proton reassignment gives a mixed-data sample after the kinematic matching that has a distribution in $\myy$ that is as statistically precise as before the matching.
 
\begin{figure}[tbp]
\centering
\includegraphics[width=0.6\textwidth]{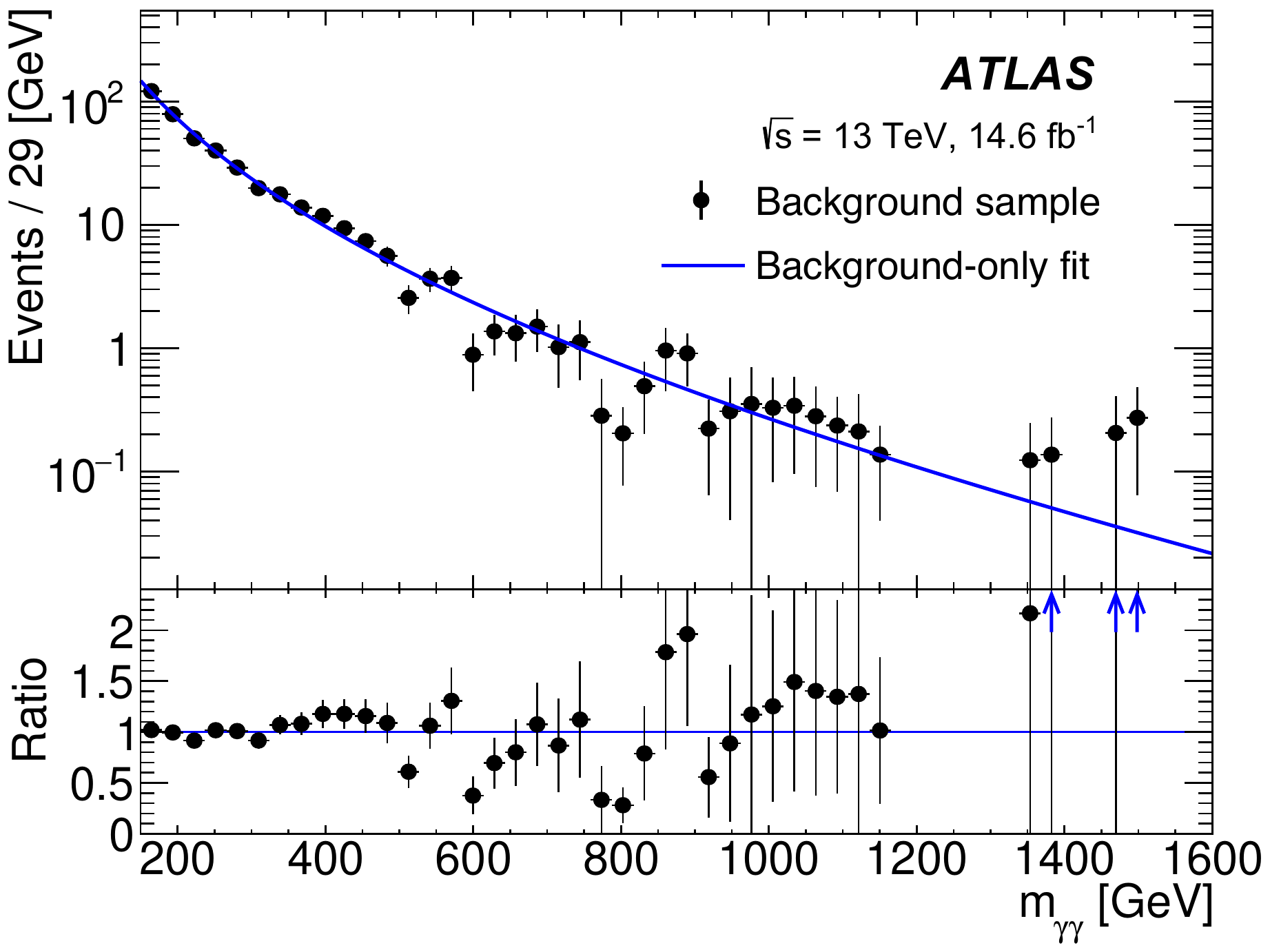}
\caption{\label{fig:bg_distribution}
Comparison of the $m_{\gamma\gamma}$ distributions after the full nominal event selection, showing the mixed-data sample (black points) and the background function fitted to it (blue line) in the diphoton mass range $[150, 1600]$~\GeV. The ratio of the mixed-data to the fitted function is shown in the lower panel.
}
\end{figure}

A fit was performed to the
mixed-data sample using the background distribution function
\begin{equation}\label{eq:bg_func}
f\left(x ; b, a_0\right)=N\left(1-x^{1 / 3}\right)^b x^{a_0},
\end{equation}
where $N$, $b$ and $a_0$ are free parameters to be fitted, and $x=m_{\gamma \gamma} / \sqrt{s}$. This function is a member of a family of functions used in previous diphoton resonance searches~\cite{HIGG-2016-08,HIGG-2018-27}.
Figure~\ref{fig:bg_distribution} shows the diphoton mass distribution of the mixed-data sample and the fitted result of the background function.
Small local deviations are in the range explained by statistical fluctuations and cannot be explained by the effects of other known systematic effects.
 
The background sample generation was validated by forming a new mixed-data sample in a different kinematic region that is orthogonal to the nominal case.
In the data and mixed-data samples, the \textit{tight} photon identification was replaced by a looser requirement~\cite{PERF-2017-01, EGAM-2018-01} and events with \textit{tight} identification were removed.
An acoplanarity requirement $\acoyy \in [0.01, 0.5]$ was applied to reduce signal contamination in the corresponding data sample. The $\myy$ distributions of this mixed-data sample and the data in the validation region after the event selection were compared, and were found to agree well in shape and normalisation.
 
Contributions from backgrounds other than the combinatorial background, collectively referred to as single-vertex backgrounds, were investigated using dedicated MC samples. Photon-induced dilepton production $\gamma\gamma\to\ell^{+}\ell^{-}$, SM light-by-light scattering $\gamma\gamma\to\gamma\gamma$, and gluon-initiated central exclusive production (CEP) were modelled at generator level with \textsc{SuperChic}\,4.13.
None of the single-vertex backgrounds has a narrow peak in the diphoton mass distribution, and their sum is small: it is below 1.2\% of the estimated combinatorial background contribution in any $m_{\gamma\gamma}$ bin of width 29~\GeV.
In addition, diffractive processes were investigated as possible single-vertex backgrounds using \Pythia[8.306]. Such contributions were also judged to be negligible. Thus, the effects of single-vertex background are neglected.
 
\begin{figure}[tbp]
\centering
\begin{tabular}{cc}
\begin{minipage}[t]{0.47\columnwidth}
\centering
\includegraphics[clip, height=0.7\textwidth]{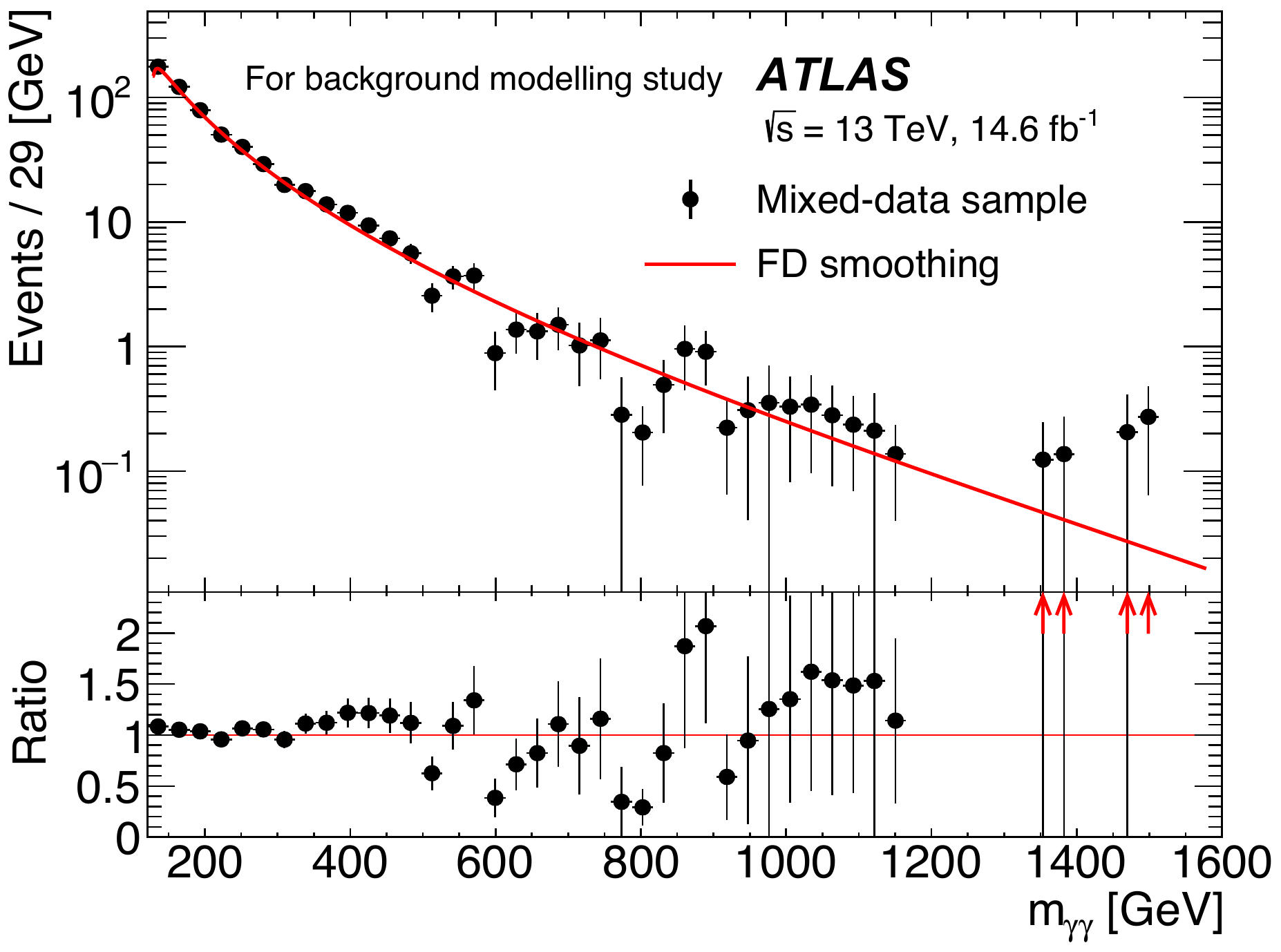}\\
(a)
\end{minipage} &
\begin{minipage}[t]{0.47\columnwidth}
\centering
\includegraphics[clip, height=0.7\textwidth]{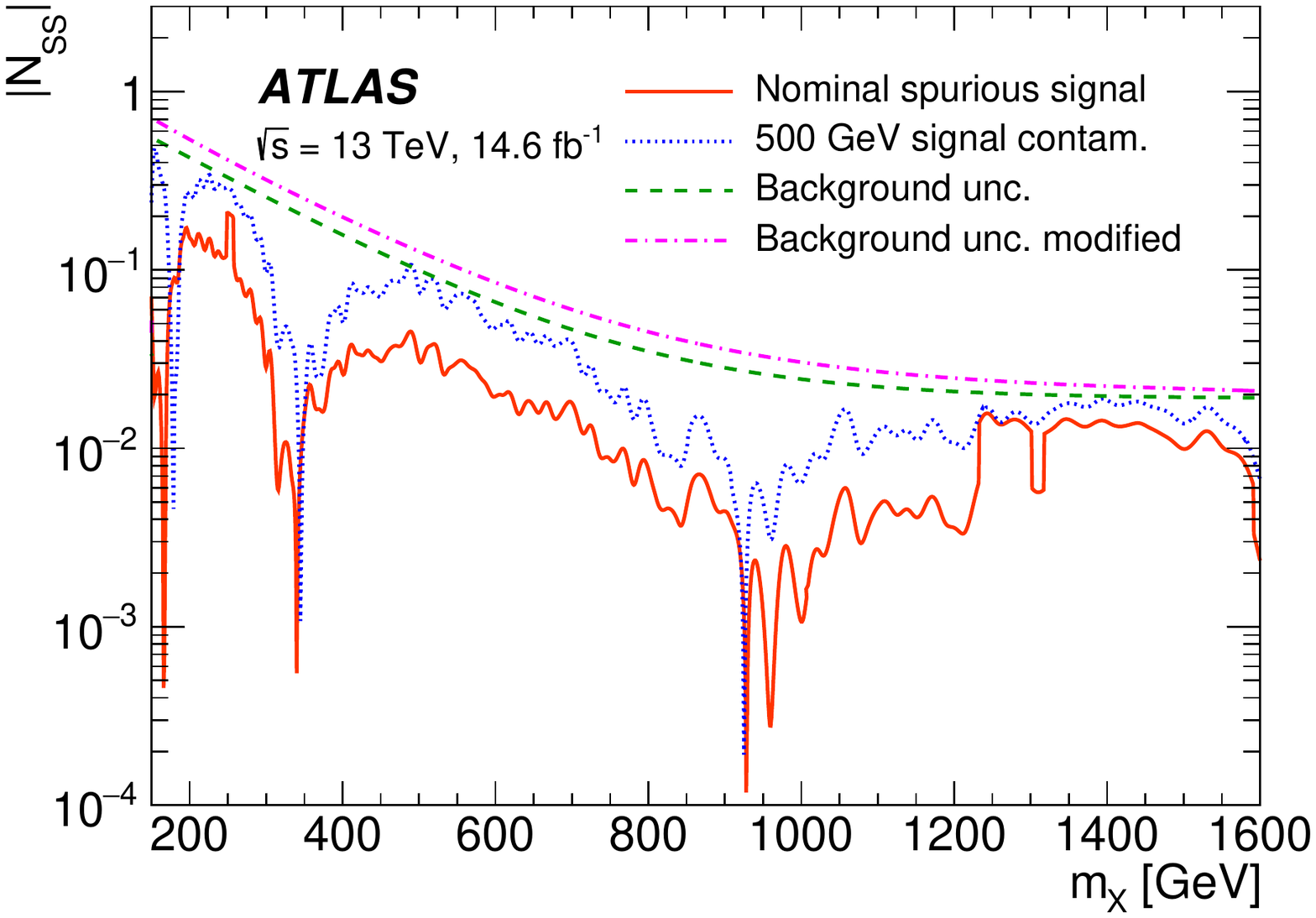}\\
(b)
\end{minipage}
\end{tabular}
\caption{\label{fig:FD_SS}
(a)~The diphoton mass distribution of the mixed-data sample (black points) and its FD smoothing result (red line). (b)~Absolute value of the number of spurious signal events $\left|N_\mathrm{SS}\right|$ as a function of hypothetical ALP mass $m_X$. The nominal (red solid line) and signal-contamination (blue dotted line) cases are illustrated. The envelope (green dashed line) is derived from the local maxima and modified considering the FD bias (magenta chain line).
}
\end{figure}

To evaluate the background-modelling uncertainty, the procedure in Ref.~\cite{HIGG-2013-08} was followed. The $\myy$ distribution of the selected mixed-data sample was smoothed using the functional decomposition (FD) method~\cite{Edgar:2018irz} as shown in Figure~\ref{fig:FD_SS}(a).
Assuming the smoothed shape to be identical to the true background distribution, signal-plus-background fits were performed for each ALP mass at 2~\GeV\ intervals in the search range.
The resulting signal yields, so-called spurious signals, are regarded as representing the degree to which the background function fails to fit the true background distribution.
The spurious signals were also evaluated for cases where ALP signals are included in the data and thus can contaminate the mixed-data sample. The ALP signals were generated as described in Section~\ref{sec:samples}, with the protons in the signal MC samples replaced by protons randomly taken from the data. The largest effect of the contamination is for $m_X=500$~\GeV. An envelope of the absolute values of the spurious signals is used as the absolute systematic uncertainty of signal yield originating from the background modelling as a function of ALP mass.
Since the assumption about the FD shape may not be exactly correct, an FD smoothing bias was evaluated to allow for the uncertainty using a procedure similar to that in Ref.~\cite{ATLAS:2022abz}.
Pseudo-datasets were generated from a number of known background shapes, taken as true background distributions. The FD smoothing result from each pseudo-dataset was compared with each shape, in terms of the spurious signals.
The relative difference in the results was combined with the original uncertainty.
Figure~\ref{fig:FD_SS}(b) shows the spurious signal for the nominal case and the largest-effect signal-contamination case. The bare envelope and the modified one considering the FD bias are also shown.
 
\section{Statistical procedure}
 
The statistical analysis uses unbinned maximum-likelihood fits made to the $m_{\gamma\gamma}$ distribution using the DSCB signal function and the background function defined in Equation~(\ref{eq:bg_func}). All the parameters of the background function are free parameters, with initial values obtained using the fit shown in figure~\ref{fig:bg_distribution}.
In these fits, the statistical uncertainty dominates over the systematic uncertainty.
The systematic uncertainties on the signal function parameters listed in Table~\ref{tab:systematics} are accounted for in the fits by using nuisance parameters constrained by Gaussian penalty terms in the likelihood function. They include the experimental uncertainties in the luminosity determination, pile-up profile, trigger efficiency, and photon identification and reconstruction.
Additional systematic uncertainties in the signal shape come from the uncertainties in the photon energy resolution and scale, but their effect on the signal yield is small.
An uncertainty in the beam optics between the central detector and AFP spectrometer affects the proton transport simulation used in proton reconstruction.
The AFP global alignment is the distance between the AFP edge and the beamline. Its preliminary measurement is corrected using the photon-initiated dimuon production events, $\gamma\gamma\to\mu^+\mu^-$~\cite{STDM-2018-16}. This correction is also regarded as the
uncertainty in the distance determination and is the largest component of the signal yield uncertainty.
The proton reconstruction efficiency and its uncertainty are determined by evaluating the efficiency for each station using the tag-and-probe method between the \textsc{Near} and \textsc{Far} stations.
Showering from the signal protons in the AFP may affect the proton reconstruction efficiency. The showering can occur in the steel window of the Roman Pot enclosing the AFP~\cite{ATLAS-TDR-18} or in the AFP itself. This effect is evaluated by varying the required number of clusters for AFP track reconstruction from two to three.
 
For the signal cross section, a theoretical uncertainty associated with the soft survival factor is included, which considers the possible presence of additional colour flow between the two scattered protons that makes both protons dissociate.
The relative uncertainty in the modelling of the soft survival factor is at the level of 1\% in the exclusive channel~\cite{Harland_Lang_2021}, with very small sensitivity to model variations in the QCD hadron-hadron interactions.
For the SD channel, the interaction is less peripheral; a reasonable variation in the modelling of soft QCD interactions gives an uncertainty of order 10\%.
Other theoretical uncertainties are neglected~\cite{PhysRevD.105.093010}.
 
\renewcommand{\arraystretch}{1.3}
\begin{table}[]
\caption{\label{tab:systematics}
Summary of the systematic uncertainties. The photon energy scale and resolution uncertainties are applied to the parameters of the DSCB function.
}
\vspace{0.3cm}
\centering
\begin{tabular}{ll}
\toprule
\textbf{Source}                                     & \textbf{Uncertainty}    \\ \midrule
\multicolumn{2}{c}{Signal yield uncertainty}                                  \\ \midrule
Pile-up reweighting                                 & $^{+2.7}_{-2.6}$\%      \\
Luminosity                                          & $\pm2.4$\%              \\
Photon identification efficiency                    & $^{+1.6}_{-1.5}$\%      \\
Photon isolation efficiency                         & $\pm1.9$\%              \\
Beam optics between ATLAS central and AFP detectors & $^{+0.8}_{-3.4}$\%      \\
AFP global alignment                                & $^{+10.0}_{-8.6}$\%     \\
Proton reconstruction efficiency                    & $^{+3.0}_{-2.2}$\%      \\
Showering in the AFP                                & $^{+0.0}_{-6.6}$\%      \\
Background modelling (mass-dependent)               & $\pm$(0.02--0.7) events \\ \midrule
\multicolumn{2}{c}{Signal modelling}                                          \\ \midrule
Photon energy resolution                            & $^{+14.1}_{-4.8}$\%     \\
Photon energy scale                                 & $\pm$(0.5--1.0)\%       \\ \midrule
\multicolumn{2}{c}{Signal cross-section uncertainty}                          \\ \midrule
Soft survival factor (exclusive process)            & $\pm2$\%                \\
Soft survival factor (single-dissociative process)  & $\pm10$\%               \\
Soft survival factor (double-dissociative process)  & $\pm50$\%               \\ \bottomrule
\end{tabular}
\end{table}
\renewcommand{\arraystretch}{1}
 
Statistical tests were performed in the search range $m_X=150$--1600~\GeV, using the conventional test statistics $q_0$ for significance and $\tilde{q}_\mu$ for exclusion upper limits~\cite{Cowan:2010js}.
The tests were performed in steps of 4~\GeV\ in $m_X$. This is smaller than the expected full resonance width, which is mostly associated with the signal mass resolution and varies from $4.3~\GeV$ for $m_X=150~\GeV$ to $27~\GeV$ for $m_X=1600~\GeV$.
Local $p$-values and their significance for the background-only hypothesis were calculated using pseudo-experiments for the $q_0$ distribution.
Global significance values were computed using the same pseudo-experiment samples to account for the look-elsewhere effect~\cite{Gross_2010,Vitells_2011}. The maximum $q_0$ with respect to $m_X$, denoted $q_0^\mathrm{max}$, was calculated for each pseudo-experiment, and the $q_0^\mathrm{max}$ distribution was used to obtain the global $p$-value.
 
Pseudo-experiments were used to derive the expected and observed 95\% confidence level (CL) exclusion upper limits on the fiducial cross section times branching ratio into two photons, computed using a modified frequentist approach, $\mathrm{CL}_\mathrm{s}$~\cite{Junk_1999,Read_2002}.

\section{Results}
 
\begin{figure}[tbp]
\centering
\includegraphics[width=0.6\textwidth]{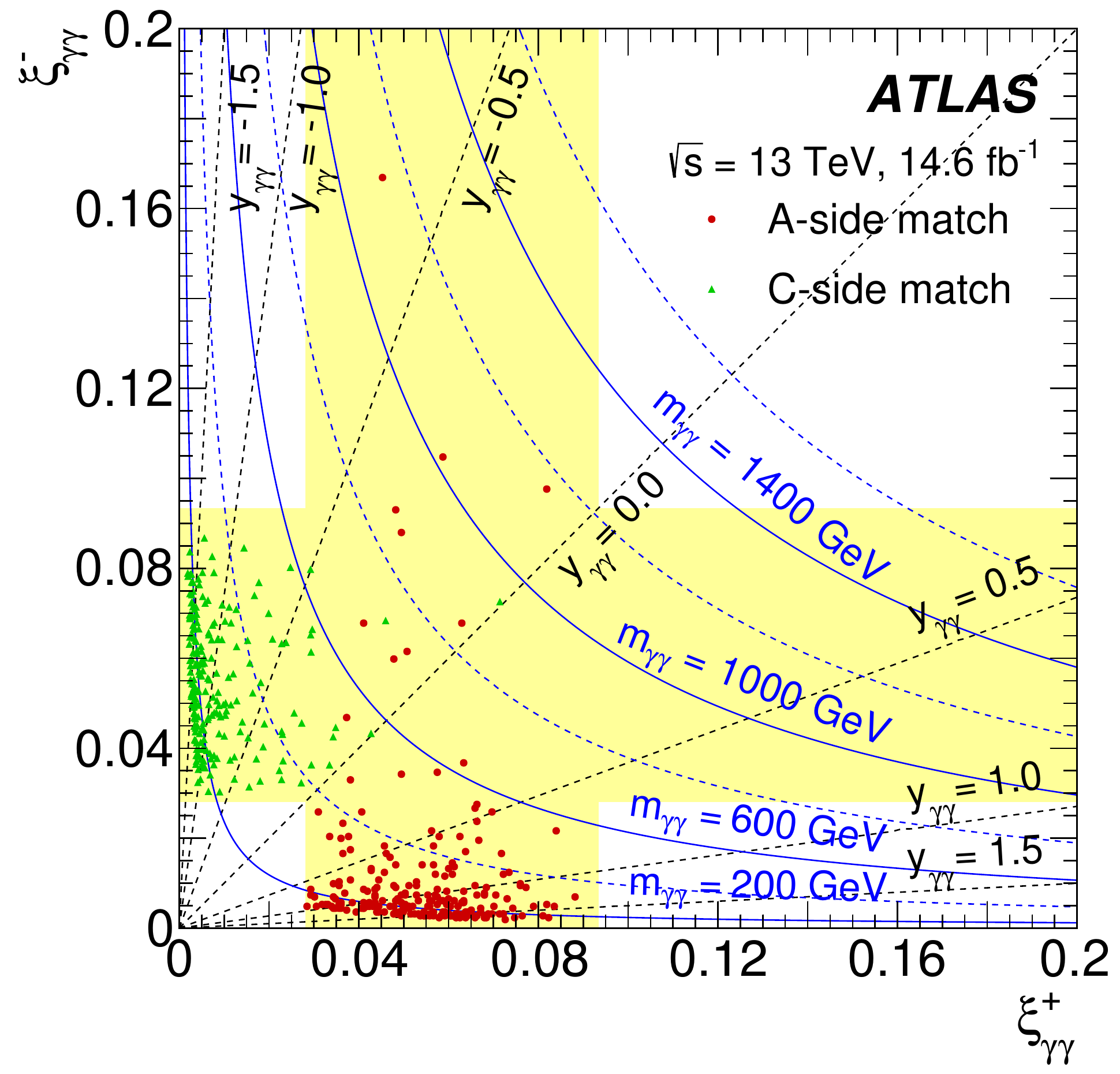}
\caption{\label{fig:xiA_xiC_data}
$(\xi_{\gamma\gamma}^+,\xi_{\gamma\gamma}^-)$ distribution of the selected data candidates after the full event selection in $m_{\gamma\gamma} \in [150,1600]$~\GeV\ with $m_{\gamma\gamma}$ contours (blue) and $y_{\gamma\gamma}$ contours (black).
The dashed blue lines are additional mass contours that bisect the spacing between adjacent solid mass contours.
The range of $\xi_{\gamma\gamma}$ in which forward-proton matching is possible, $[0.035-\xi_\mathrm{th}, 0.08+\xi_\mathrm{th}]$, is indicated by the yellow rectangle for each side. Events passing the matching requirement on the A(C)-side are represented by the red dots (green triangles). No event passed the matching requirement for both the A-side and C-side.
}
\end{figure}

\begin{figure}[tbp]
\centering
\includegraphics[width=0.6\textwidth]{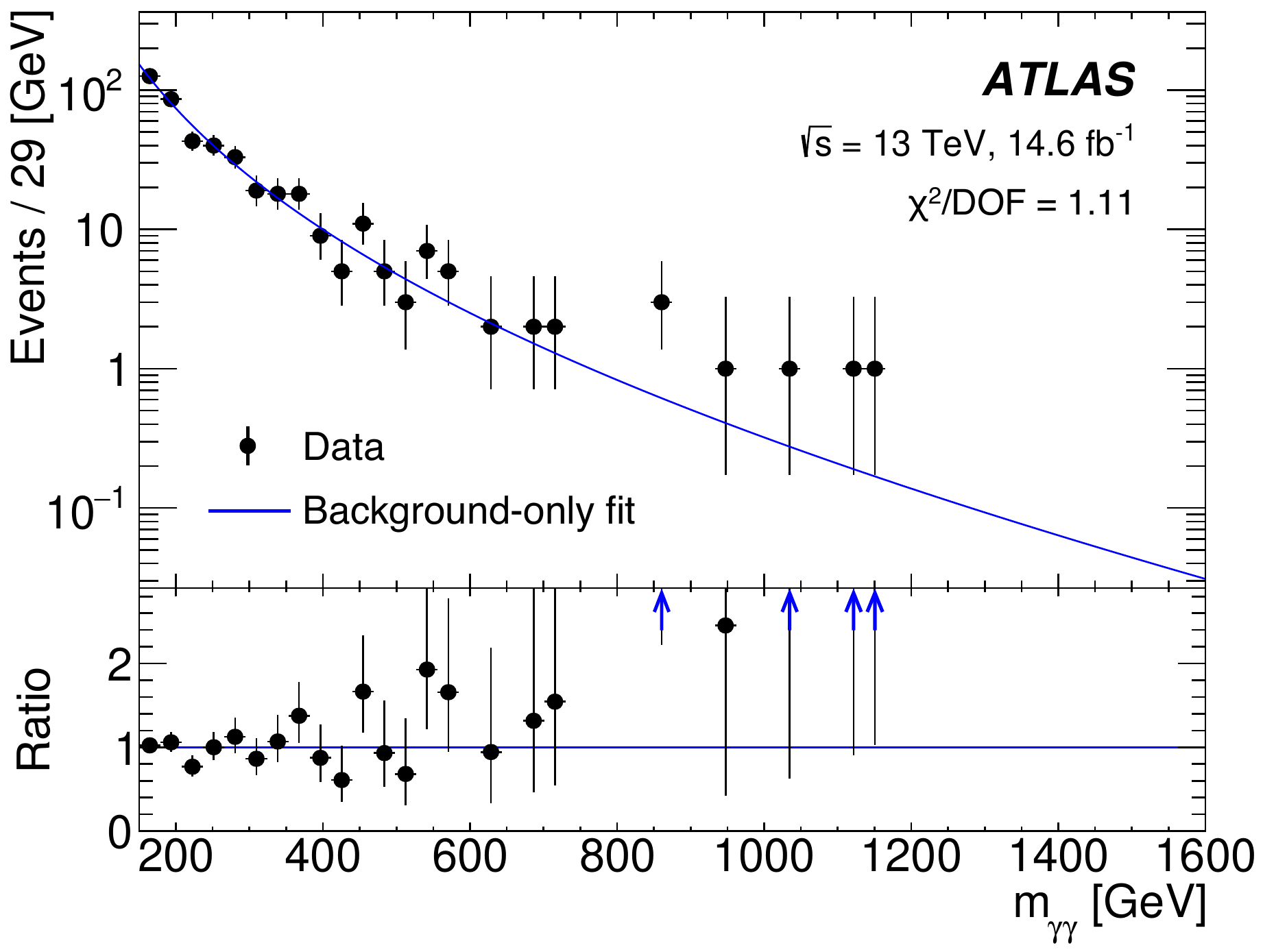}
\caption{\label{fig:bfit_result_data}
Background-only fit to the observed numbers of events (black points) as a function of the diphoton invariant mass $m_{\gamma\gamma}$. The ratio of the data points to the fitted background (blue line) is shown in the bottom panel.
The quantity $\chi^2$/DOF is the chi-squared per degree of freedom in the fit to data.
}
\end{figure}

\begin{figure}[tbp]
\centering
\includegraphics[width=0.6\textwidth]{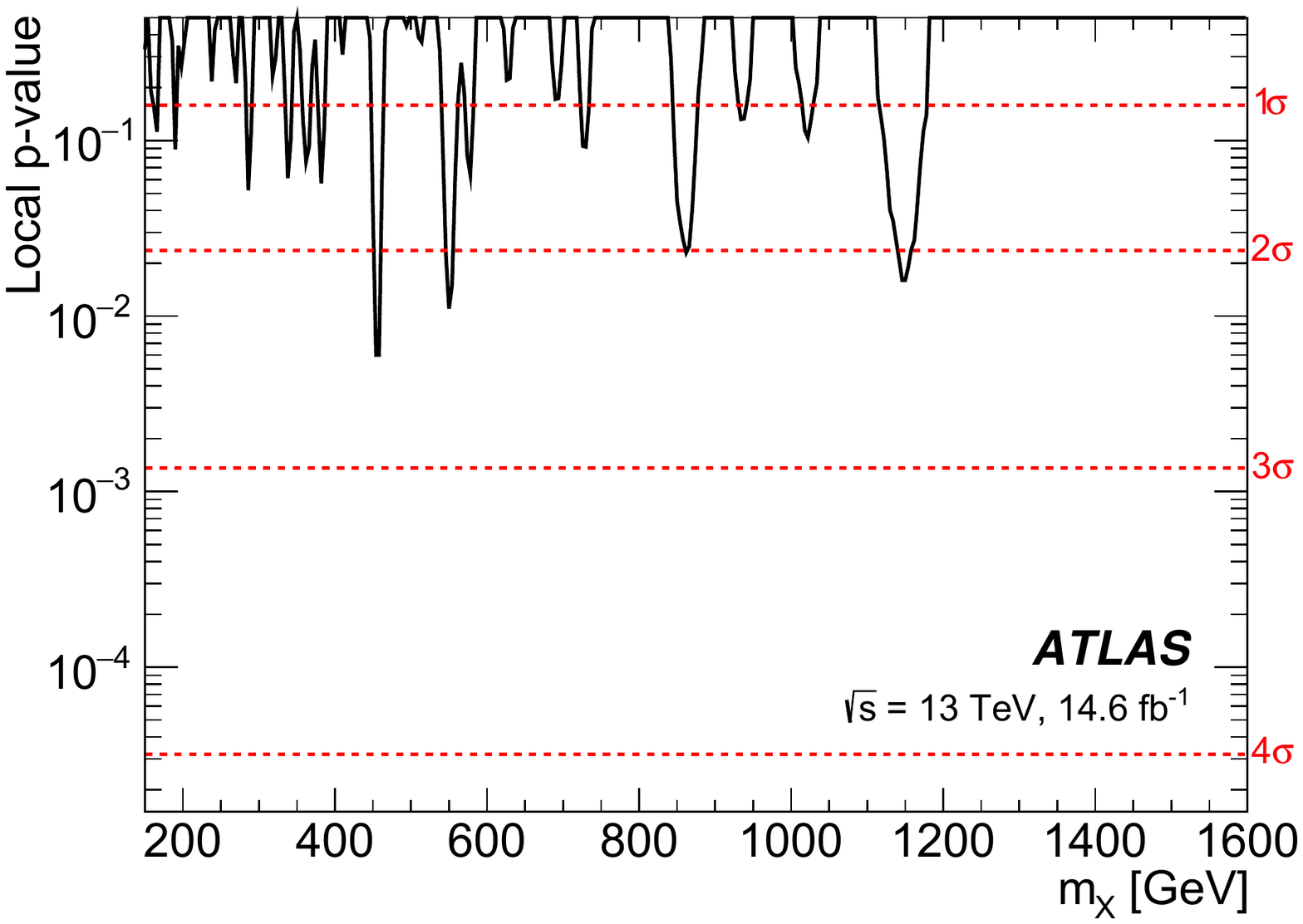}
\caption{\label{fig:unblinded_data_fit_p_Z}
Local $p$-values for ALP observation on the basis of a null hypothesis as a function of $m_X$. Significances corresponding to $1\sigma$, $2\sigma$, $3\sigma$, and $4\sigma$ in units of standard deviation are indicated by the red dashed lines.
}
\end{figure}

A total of 441 events are observed in the range $m_{\gamma\gamma}\in[150,1600]$~\GeV. Figure~\ref{fig:xiA_xiC_data} shows $\xi_{\gamma\gamma}^+$ vs $\xi_{\gamma\gamma}^-$ for these events with $m_{\gamma\gamma}$ and $y_{\gamma\gamma}$ contours.
In this sample, 219 (222) events passed the matching selection for the A(C)-side; no event passed the matching for both the A-side and C-side, in accordance with expectations. The diphoton invariant-mass distribution for the selected events is shown in Figure~\ref{fig:bfit_result_data}, along with the background-only fit. The highest-mass diphoton candidate observed in the data has a mass of 1.16~\TeV.
The probability of compatibility with the background-only hypothesis, quantified as the local $p$-value, is shown in Figure~\ref{fig:unblinded_data_fit_p_Z} as a function of the hypothesised ALP resonance mass. The most significant excess, observed at $m_X=454$~\GeV, has a local significance of $2.51\sigma$. The global $p$-value for the null hypothesis is larger than 0.5, from which it is concluded that no significant excess over the background-only hypothesis is observed.
 
\begin{figure}[tbp]
\centering
\begin{tabular}{cc}
\begin{minipage}[t]{0.47\columnwidth}
\centering
\includegraphics[clip, width=\textwidth]{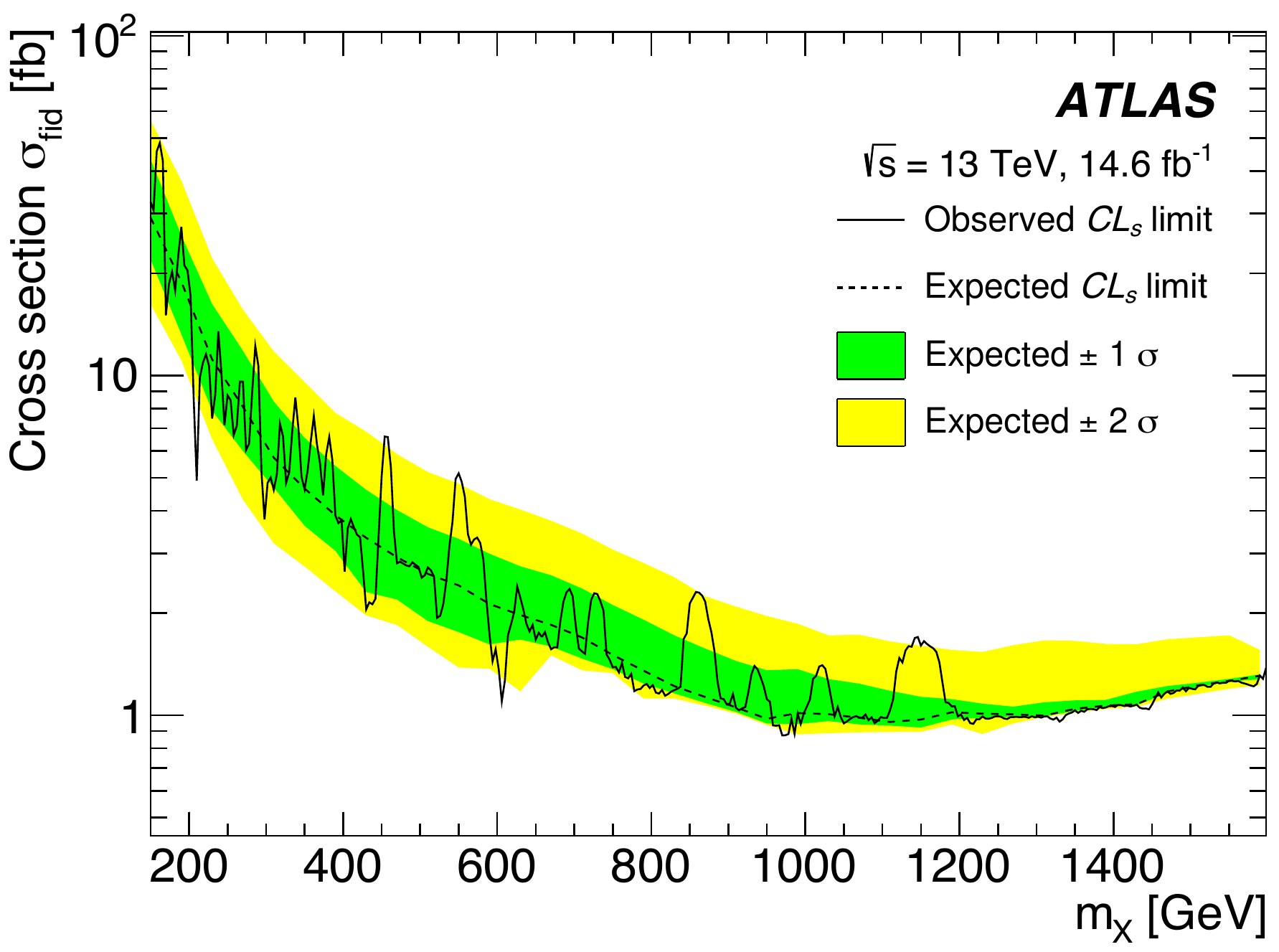}\\
(a)
\end{minipage} &
\begin{minipage}[t]{0.47\columnwidth}
\centering
\includegraphics[clip, width=\textwidth]{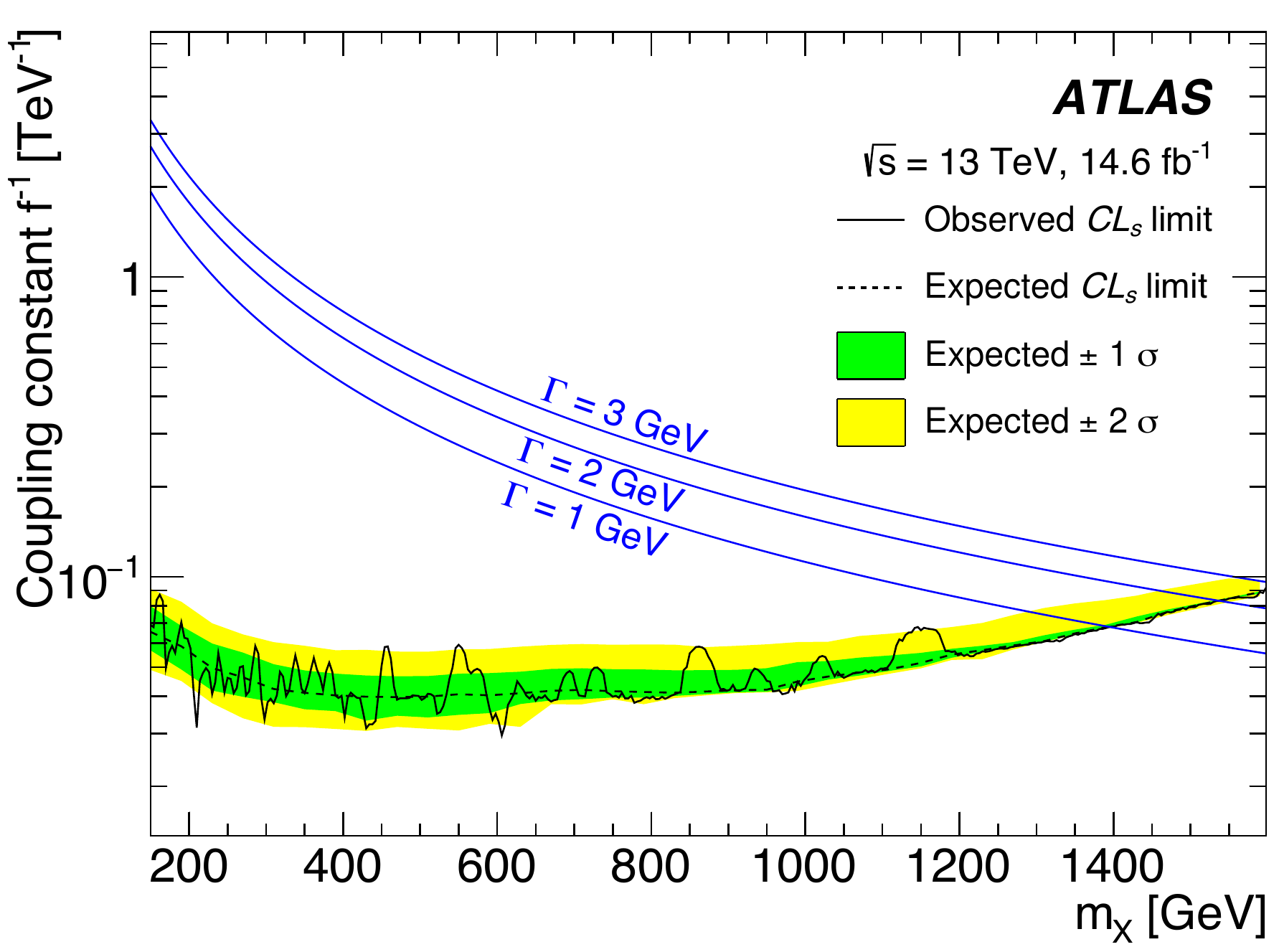}\\
(b)
\end{minipage}
\end{tabular}
\caption{\label{fig:limit}
Expected and observed  95\% CL upper limits on (a)~the signal fiducial cross section $\sigma_\mathrm{fid}$ and (b)~the ALP coupling constant, assuming 100\% branching ratio for ALP decay into two photons, as functions of the hypothetical ALP mass $m_X$. The $1\sigma$ and $2\sigma$ confidence intervals are shown by the coloured bands. Contours of the ALP natural width $\Gamma$ are illustrated by the smooth blue solid lines.
}
\end{figure}

The observed and expected upper limits on the fiducial cross section and coupling constant for ALP production at 95\% confidence level, assuming a 100\% decay branching ratio into two photons, are shown in Figure~\ref{fig:limit} and are in good agreement, consistent with the absence of a signal.  
The upper limits on the coupling constant are derived using the relationship between the expected fiducial cross section and coupling constant in Ref.~\cite{Baldenegro:2018hng}.
The observed limit on the coupling constant is in the range 0.04--0.09~$\TeV^{-1}$.
The ALP natural width is $\Gamma\approx 1$~\GeV\ for $m_X=1400$~\GeV\ and $\Gamma\approx 3$~\GeV\ for $m_X=1600$~\GeV\ at the observed limit. Such widths are sufficiently small to be ignored, relative to the detector resolution. The exclusion of a coupling constant $f^{-1}$ much larger than illustrated by the $\Gamma$ contours is not justified in this analysis because it violates the narrow-width approximation used in the signal modelling.
 
A previous inclusive diphoton resonance search~\cite{HIGG-2018-27} also has sensitivity to the ALP signal. To estimate the sensitivity of that search to the signal considered in this paper, the efficiency of the ALP signal to pass the event selections defined in that paper, relative to the corresponding fiducial requirements, was evaluated and was found to be similar to the other production mechanisms considered. The inclusive analysis is sensitive to smaller ALP production cross-sections, partly because of the higher integrated luminosity and partly because of the effect of the restricted $\xi$ range used in this paper. However, by requiring a forward-proton tag, the present approach is able to add valuable information and can distinguish cleanly a photon-induced production mechanism from other production mechanisms, should the presence of a signal be observed.
The present analysis demonstrates the applicability of forward proton tagging to add precision to ALP searches that may be carried out using present and future data samples.  In particular, the sufficiency of a single AFP tag has been shown, along with a method for optimising the background calculation.  Both these features validate and improve the statistical significance of the forward-tagging approach.
 
\FloatBarrier
 
\section{Conclusion}
 
A search for an axion-like particle (ALP) has been carried out with the ATLAS experiment
using 14.6~fb$^{-1}$ of $\sqrt{s} = 13~\TeV$ $pp$ collisions at the LHC.
Events with centrally produced photon pairs tagged by forward scattered protons have been studied in a search for light-by-light scattering, $pp \to p(\gamma\gamma\to\gamma\gamma)p^{(*)}$, mediated by an ALP resonance.  The search was performed in the diphoton mass range $m_{\gamma\gamma}=[150,1600]$~\GeV.
No signal is observed, and the data are consistent with a smooth combinatorial background that can be assumed to come from Standard Model processes.
The inferred upper limit on the ALP coupling constant, assuming a 100\% decay branching ratio into two photons, is in the range 0.04--0.09~$\TeV^{-1}$
at 95\% confidence level.
These results are comparable to those of the CMS-TOTEM collaboration, obtained using a similar approach, and extend their measured mass range to lower values.
 
\section*{Acknowledgements}

% The next lines are included from the .//acknowledgements/Acknowledgements.tex input file

We thank CERN for the very successful operation of the LHC, as well as the
support staff from our institutions without whom ATLAS could not be
operated efficiently.
 
We acknowledge the support of
ANPCyT, Argentina;
YerPhI, Armenia;
ARC, Australia;
BMWFW and FWF, Austria;
ANAS, Azerbaijan;
CNPq and FAPESP, Brazil;
NSERC, NRC and CFI, Canada;
CERN;
ANID, Chile;
CAS, MOST and NSFC, China;
Minciencias, Colombia;
MEYS CR, Czech Republic;
DNRF and DNSRC, Denmark;
IN2P3-CNRS and CEA-DRF/IRFU, France;
SRNSFG, Georgia;
BMBF, HGF and MPG, Germany;
GSRI, Greece;
RGC and Hong Kong SAR, China;
ISF and Benoziyo Center, Israel;
INFN, Italy;
MEXT and JSPS, Japan;
CNRST, Morocco;
NWO, Netherlands;
RCN, Norway;
MEiN, Poland;
FCT, Portugal;
MNE/IFA, Romania;
MESTD, Serbia;
MSSR, Slovakia;
ARRS and MIZ\v{S}, Slovenia;
DSI/NRF, South Africa;
MICINN, Spain;
SRC and Wallenberg Foundation, Sweden;
SERI, SNSF and Cantons of Bern and Geneva, Switzerland;
MOST, Taiwan;
TENMAK, T\"urkiye;
STFC, United Kingdom;
DOE and NSF, United States of America.
In addition, individual groups and members have received support from
BCKDF, CANARIE, Compute Canada and CRC, Canada;
PRIMUS 21/SCI/017 and UNCE SCI/013, Czech Republic;
COST, ERC, ERDF, Horizon 2020 and Marie Sk{\l}odowska-Curie Actions, European Union;
Investissements d'Avenir Labex, Investissements d'Avenir Idex and ANR, France;
DFG and AvH Foundation, Germany;
Herakleitos, Thales and Aristeia programmes co-financed by EU-ESF and the Greek NSRF, Greece;
BSF-NSF and MINERVA, Israel;
Norwegian Financial Mechanism 2014-2021, Norway;
NCN and NAWA, Poland;
La Caixa Banking Foundation, CERCA Programme Generalitat de Catalunya and PROMETEO and GenT Programmes Generalitat Valenciana, Spain;
G\"{o}ran Gustafssons Stiftelse, Sweden;
The Royal Society and Leverhulme Trust, United Kingdom.
 
The crucial computing support from all WLCG partners is acknowledged gratefully, in particular from CERN, the ATLAS Tier-1 facilities at TRIUMF (Canada), NDGF (Denmark, Norway, Sweden), CC-IN2P3 (France), KIT/GridKA (Germany), INFN-CNAF (Italy), NL-T1 (Netherlands), PIC (Spain), ASGC (Taiwan), RAL (UK) and BNL (USA), the Tier-2 facilities worldwide and large non-WLCG resource providers. Major contributors of computing resources are listed in Ref.~\cite{ATL-SOFT-PUB-2021-003}.

% End of text imported from the .//acknowledgements/Acknowledgements.tex input file

\printbibliography

@article{Knapen_2017,
	doi = {10.1103/physrevlett.118.171801},
	year = 2017,
	volume = {118},
	number = {17},
       pages          = "171801",
	author = {Simon Knapen and Tongyan Lin and Hou Keong Lou and Tom Melia},
	title = {Searching for Axionlike Particles with Ultraperipheral Heavy-Ion Collisions},
	journal = {Phys. Rev. Lett.},
 eprint         = "1607.06083",
      archivePrefix  = "arXiv",
      primaryClass   = "hep-ph",
}

@article{Baldenegro:2018hng,
  author        = {Baldenegro, Cristian and Fichet, Sylvain and von Gersdorff, Gero and Royon, Christophe},
  title         = {{Searching for axion-like particles with proton tagging at the LHC}},
  journal       = {JHEP},
  volume        = {06},
  year          = {2018},
  pages         = {131},
  doi           = {10.1007/JHEP06(2018)131},
  eprint        = {1803.10835},
  archiveprefix = {arXiv},
  primaryclass  = {hep-ph}
}

@article{Lange:2016dro,
    author = "Lange, J. and others",
    collaboration = "ATLAS",
    title = "{Beam tests of an integrated prototype of the ATLAS Forward Proton detector}",
    eprint = "1608.01485",
    archivePrefix = "arXiv",
    primaryClass = "physics.ins-det",
    doi = "10.1088/1748-0221/11/09/P09005",
    journal = "JINST",
    volume = "11",
    number = "09",
    pages = "P09005",
    year = "2016"
}

@article{Lange:2015xla,
    author = "Lange, J. and Cavallaro, E. and Grinstein, S. and López Paz, I.",
    title = "{3D silicon pixel detectors for the ATLAS Forward Physics experiment}",
    eprint = "1501.02076",
    archivePrefix = "arXiv",
    primaryClass = "physics.ins-det",
    doi = "10.1088/1748-0221/10/03/C03031",
    journal = "JINST",
    volume = "10",
    number = "03",
    pages = "C03031",
    year = "2015"
}

@article{Grinstein:2016hmq,
    author = "Grinstein, S. and others",
    title = "{Module production of the one-arm AFP 3D pixel tracker}",
    eprint = "1611.01005",
    archivePrefix = "arXiv",
    primaryClass = "physics.ins-det",
    doi = "10.1088/1748-0221/12/01/C01086",
    journal = "JINST",
    volume = "12",
    number = "01",
    pages = "C01086",
    year = "2017"
}

@article{GarciaSciveres:2011zz,
    author = "Garcia-Sciveres, M. and others",
    title = "{The FE-I4 pixel readout integrated circuit}",
    doi = "10.1016/j.nima.2010.04.101",
    journal = "Nucl. Instrum. Meth. A",
    volume = "636",
    pages = "S155--S159",
    year = "2011"
}

@article{Zivkovic_2012,
  doi = {10.1088/1748-0221/7/02/c02050},
  year = 2012,
  month = {feb},
  publisher = {{IOP} Publishing},
  volume = {7},
  number = {02},
  pages = {C02050--C02050},
  author = {V Zivkovic and others},
  title = {The {FE}-I4 pixel readout system-on-chip resubmission for the insertable B-Layer project},
  journal = {JINST},
  
}

@article{Kocian:2017ygu,
    author = "Kocian, M.",
    collaboration = "ATLAS",
    title = "{Readout and trigger for the AFP detector at ATLAS experiment}",
    reportNumber = "ATL-DAQ-PROC-2016-031",
    doi = "10.1088/1748-0221/12/01/C01077",
    journal = "JINST",
    volume = "12",
    number = "01",
    pages = "C01077",
    year = "2017"
}

@article{10.21468/SciPostPhysCodeb.8,
  title         = {{A comprehensive guide to the physics and usage of PYTHIA 8.3}},
  author        = {Christian Bierlich and Smita Chakraborty and Nishita Desai and Leif Gellersen and Ilkka Helenius and Philip Ilten and Leif Lönnblad and Stephen Mrenna and Stefan Prestel and Christian T. Preuss and Torbjörn Sjöstrand and Peter Skands and Marius Utheim and Rob Verheyen},
  journal       = {SciPost Phys. Codebases},
  pages         = {8},
  year          = {2022},
  doi           = {10.21468/SciPostPhysCodeb.8},
  eprint        = {2203.11601},
  archiveprefix = {arXiv},
  primaryclass  = {hep-ph}
}

@article{Edgar:2018irz,
    author        = {Edgar, Ryan and Amidei, Dante and Grud, Christopher and Sekhon, Karishma},
  title         = {{Functional Decomposition: A new method for search and limit setting}},
  eprint        = {1805.04536},
  archiveprefix = {arXiv},
  primaryclass  = {physics.data-an},
  year          = {2018}
}

@article{PhysRevSTAB.15.051002,
  title = {Semiautomatic beam-based LHC collimator alignment},
  author = {Valentino, Gianluca and A\ss{}mann, Ralph and Bruce, Roderik and Redaelli, Stefano and Rossi, Adriana and Sammut, Nicholas and Wollmann, Daniel},
  journal = {Phys. Rev. ST Accel. Beams},
  volume = {15},
  pages = {051002},
  numpages = {12},
  year = {2012},
  month = {May},
  publisher = {American Physical Society},
  doi = {10.1103/PhysRevSTAB.15.051002},
}

@article{Zamantzas2007,
  author={C. {Zamantzas} and B. {Dehning} and E. {Effinger} and J. {Emery} and G. {Ferioli} and S. {Jackson}},
  journal={IEEE Nucl. Sci. Symp. Conf. Record}, 
  title={The LHC beam loss monitoring system’s data contribution to other systems}, 
  year={2007},
  volume={3},
  number={},
  pages={2331-2335},
  doi={https://doi.org/10.1109/NSSMIC.2007.4436612}
}

@article{PhysRevAccelBeams.20.081002,
  title = {Final implementation, commissioning, and performance of embedded collimator beam position monitors in the Large Hadron Collider},
  author = {Valentino, Gianluca and Baud, Guillaume and Bruce, Roderik and Gasior, Marek and Mereghetti, Alessio and Mirarchi, Daniele and Olexa, Jakub and Redaelli, Stefano and Salvachua, Belen and Valloni, Alessandra and Wenninger, Jorg},
  journal = {Phys. Rev. Accel. Beams},
  volume = {20},
  pages = {081002},
  numpages = {13},
  year = {2017},
  month = {Aug},
  publisher = {American Physical Society},
  doi = {10.1103/PhysRevAccelBeams.20.081002},
}

@article{sc2_ref,
  doi           = {10.1140/epjc/s10052-015-3832-8},
  year          = 2016,
  volume        = {76},
  number        = {1},
  pages         = {9},
  author        = {L. A. Harland-Lang and V. A. Khoze and M. G. Ryskin},
  title         = {Exclusive physics at the {LHC} with {SuperChic} 2},
  journal       = {Eur. Phys. J. C},
  eprint        = {1508.02718},
  archiveprefix = {arXiv},
  primaryclass  = {hep-ph}
}

@article{sc3_ref,
  doi           = {10.1140/epjc/s10052-018-6530-5},
  year          = 2019,
  volume        = {79},
  number        = {1},
  pages         = {39},
  author        = {L. A. Harland-Lang and V. A. Khoze and M. G. Ryskin},
  title         = {Exclusive {LHC} physics with heavy ions: {SuperChic} 3},
  journal       = {Eur. Phys. J. C},
  eprint        = {1810.06567},
  archiveprefix = {arXiv},
  primaryclass  = {hep-ph}
}

@article{sc4_ref,
  doi           = {10.1140/epjc/s10052-020-08455-0},
  year          = 2020,
  volume        = {80},
  number        = {10},
  pages         = {925},
  author        = {L. A. Harland-Lang and M. Tasevsky and V. A. Khoze and M. G. Ryskin},
  title         = {A new approach to modelling elastic and inelastic photon-initiated production at the {LHC}: {SuperChic} 4},
  journal       = {Eur. Phys. J. C},
  eprint        = {2007.12704},
  archiveprefix = {arXiv},
  primaryclass  = {hep-ph}
}

@article{sc4_alpSDDD,
  author = "Harland-Lang, L. A. and Tasevsky, M.",
   title = "{Searching for axionlike particles with proton tagging and rapidity gaps at the LHC}",
   doi = "10.1103/PhysRevD.107.033001",
   journal = "Phys. Rev. D",
   volume = "107",
   number = "3",
   pages = "033001",
   year = "2023",
  eprint        = {2208.10526},
  archiveprefix = {arXiv},
  primaryclass  = {hep-ph}
}

@report{Oreglia:1980cs,
  author = {Oreglia, M.},
  title  = {{A Study of the Reactions $\psi^\prime \to \gamma \gamma \psi$}},
  type   = {SLAC-R-236},
  year   = {1980}
}

@booklet{ATLAS:2022abz,
  author        = {{ATLAS Collaboration}},
  title         = {{Search for boosted diphoton resonances in the 10 to 70 GeV mass range using 138 fb$^{-1}$ of 13 TeV $pp$ collisions with the ATLAS detector}},
  eprint        = {2211.04172},
  archiveprefix = {arXiv},
  primaryclass  = {hep-ex},
  howpublished  = {{CERN-EP-2022-183}},
  year          = {2022}
}

@article{Read_2002,
  author  = {Read, Alexander L.},
  journal = {J. Phys. G},
  pages   = {2693},
  doi     = {10.1088/0954-3899/28/10/313},
  title   = {{Presentation of search results: the \(CL_S\) technique}},
  volume  = {28},
  year    = {2002}
}

@article{Junk_1999,
  doi           = {10.1016/s0168-9002(99)00498-2},
  year          = 1999,
  volume        = {434},
  number        = {2-3},
  pages         = {435--443},
  author        = {Thomas Junk},
  title         = {Confidence level computation for combining searches with small statistics},
  journal       = {Nucl. Instrum. Meth. Phys. Res. A},
  eprint        = {hep-ex/9902006},
  archiveprefix = {arXiv}
}

@article{Gross_2010,
  doi           = {10.1140/epjc/s10052-010-1470-8},
  year          = 2010,
  volume        = {70},
  number        = {1-2},
  pages         = {525--530},
  author        = {Eilam Gross and Ofer Vitells},
  title         = {Trial factors for the look elsewhere effect in high energy physics},
  journal       = {Eur. Phys. J. C},
  eprint        = {1005.1891},
  archiveprefix = {arXiv},
  primaryclass  = {physics.data-an}
}

@article{Vitells_2011,
  doi           = {10.1016/j.astropartphys.2011.08.005},
  year          = 2011,
  volume        = {35},
  number        = {5},
  pages         = {230--234},
  author        = {Ofer Vitells and Eilam Gross},
  title         = {Estimating the significance of a signal in a multi-dimensional search},
  journal       = {Astropart. Phys.},
  eprint        = {1105.4355},
  archiveprefix = {arXiv},
  primaryclass  = {astro-ph.IM}
}

@article{PhysRevLett.97.131801,
  title         = {Improved Experimental Limit on the Electric Dipole Moment of the Neutron},
  author        = {Baker, C. A. and Doyle, D. D. and Geltenbort, P. and Green, K. and van der Grinten, M. G. D. and Harris, P. G. and Iaydjiev, P. and Ivanov, S. N. and May, D. J. R. and Pendlebury, J. M. and Richardson, J. D. and Shiers, D. and Smith, K. F.},
  journal       = {Phys. Rev. Lett.},
  volume        = {97},
  pages         = {131801},
  numpages      = {4},
  year          = {2006},
  doi           = {10.1103/PhysRevLett.97.131801},
  eprint        = {hep-ex/0602020},
  archiveprefix = {arXiv}
}

@article{PhysRevLett.124.081803,
  title         = {Measurement of the Permanent Electric Dipole Moment of the Neutron},
  author        = {Abel, C. and Afach, S. and Ayres, N. J. and Baker, C. A. and Ban, G. and Bison, G. and Bodek, K. and Bondar, V. and Burghoff, M. and Chanel, E. and Chowdhuri, Z. and Chiu, P.-J. and Clement, B. and Crawford, C. B. and Daum, M. and Emmenegger, S. and Ferraris-Bouchez, L. and Fertl, M. and Flaux, P. and Franke, B. and Fratangelo, A. and Geltenbort, P. and Green, K. and Griffith, W. C. and van der Grinten, M. and Gruji\ifmmode \acute{c}\else \'{c}\fi{}, Z. D. and Harris, P. G. and Hayen, L. and Heil, W. and Henneck, R. and H\'elaine, V. and Hild, N. and Hodge, Z. and Horras, M. and Iaydjiev, P. and Ivanov, S. N. and Kasprzak, M. and Kermaidic, Y. and Kirch, K. and Knecht, A. and Knowles, P. and Koch, H.-C. and Koss, P. A. and Komposch, S. and Kozela, A. and Kraft, A. and Krempel, J. and Ku\ifmmode \acute{z}\else \'{z}\fi{}niak, M. and Lauss, B. and Lefort, T. and Lemi\`ere, Y. and Leredde, A. and Mohanmurthy, P. and Mtchedlishvili, A. and Musgrave, M. and Naviliat-Cuncic, O. and Pais, D. and Piegsa, F. M. and Pierre, E. and Pignol, G. and Plonka-Spehr, C. and Prashanth, P. N. and Qu\'em\'ener, G. and Rawlik, M. and Rebreyend, D. and Rien\"acker, I. and Ries, D. and Roccia, S. and Rogel, G. and Rozpedzik, D. and Schnabel, A. and Schmidt-Wellenburg, P. and Severijns, N. and Shiers, D. and Tavakoli Dinani, R. and Thorne, J. A. and Virot, R. and Voigt, J. and Weis, A. and Wursten, E. and Wyszynski, G. and Zejma, J. and Zenner, J. and Zsigmond, G.},
  journal       = {Phys. Rev. Lett.},
  volume        = {124},
  pages         = {081803},
  numpages      = {7},
  year          = {2020},
  doi           = {10.1103/PhysRevLett.124.081803},
  eprint        = {2001.11966},
  archiveprefix = {arXiv},
  primaryclass  = {hep-ex}
}

@article{PhysRevLett.38.1440,
  title    = {$\mathrm{CP}$ Conservation in the Presence of Pseudoparticles},
  author   = {Peccei, R. D. and Quinn, Helen R.},
  journal  = {Phys. Rev. Lett.},
  volume   = {38},
  pages    = {1440--1443},
  numpages = {0},
  year     = {1977},
  doi      = {10.1103/PhysRevLett.38.1440}
}

@article{PhysRevD.16.1791,
  title    = {Constraints imposed by $\mathrm{CP}$ conservation in the presence of pseudoparticles},
  author   = {Peccei, R. D. and Quinn, Helen R.},
  journal  = {Phys. Rev. D},
  volume   = {16},
  pages    = {1791--1797},
  numpages = {0},
  year     = {1977},
  doi      = {10.1103/PhysRevD.16.1791}
}

@article{PhysRevLett.40.223,
  title    = {A New Light Boson?},
  author   = {Weinberg, Steven},
  journal  = {Phys. Rev. Lett.},
  volume   = {40},
  pages    = {223--226},
  numpages = {0},
  year     = {1978},
  doi      = {10.1103/PhysRevLett.40.223}
}

@article{PhysRevLett.40.279,
  title    = {Problem of Strong $P$ and $T$ Invariance in the Presence of Instantons},
  author   = {Wilczek, F.},
  journal  = {Phys. Rev. Lett.},
  volume   = {40},
  pages    = {279--282},
  numpages = {0},
  year     = {1978},
  doi      = {10.1103/PhysRevLett.40.279}
}

@article{Staszewski_2009,
	doi = {10.1016/j.nima.2009.08.023},
	year = 2009,
	volume = {609},
	number = {2-3},
	pages = {136--141},
	author = {R. Staszewski and J. Chwastowski},
	title = {Transport simulation and diffractive event reconstruction at the {LHC}},
	journal = {Nucl. Instrum. Meth. A},
 eprint        = {0906.2868},
  archiveprefix = {arXiv},
  primaryclass  = {physics.ins-det}
}

@article{Bruce_2020,
  doi           = {10.1088/1361-6471/ab7ff7},
  year          = {2020},
  publisher     = {IOP Publishing},
  volume        = {47},
  number        = {6},
  pages         = {060501},
  author        = {Roderik Bruce and David d'Enterria and Albert de Roeck and Marco Drewes and Glennys R Farrar and Andrea Giammanco and Oliver Gould and Jan Hajer and Lucian Harland-Lang and Jan Heisig and John M Jowett and Sonia Kabana and Georgios K Krintiras and Michael Korsmeier and Michele Lucente and Guilherme Milhano and Swagata Mukherjee and Jeremi Niedziela and Vitalii A Okorokov and Arttu Rajantie and Michaela Schaumann},
  title         = {New physics searches with heavy-ion collisions at the CERN Large Hadron Collider},
  journal       = {J. Phys. G},
  eprint        = {1812.07688},
  archiveprefix = {arXiv},
  primaryclass  = {hep-ph}
}

@article{Cacciari_2008,
	doi = {10.1088/1126-6708/2008/04/005},
	year = 2008,
	volume = {04},
	number = {04},
	pages = {005--005},
	author = {Matteo Cacciari and Gavin P Salam and Gregory Soyez},
	title = {The catchment area of jets},
	journal = {JHEP},
 eprint        = {0802.1188},
  archiveprefix = {arXiv},
  primaryclass  = {hep-ph}
}

@article{Cacciari_2010,
	doi = {10.1007/jhep04(2010)065},
	year = 2010,
	volume = {04},
	number = {4},
        pages = {065},
	author = {Matteo Cacciari and Gavin P. Salam and Sebastian Sapeta},
	title = {On the characterisation of the underlying event},
	journal = {JHEP},
eprint        = {0912.4926},
  archiveprefix = {arXiv},
  primaryclass  = {hep-ph}
}

@article{PhysRevD.105.093010,
  title         = {Modeling ${W}^{+}{W}^{\ensuremath{-}}$ production with rapidity gaps at the LHC},
  author        = {Bailey, S. and Harland-Lang, L. A.},
  journal       = {Phys. Rev. D},
  volume        = {105},
  pages         = {093010},
  numpages      = {32},
  year          = {2022},
  doi           = {10.1103/PhysRevD.105.093010},
  eprint        = {2201.08403},
  archiveprefix = {arXiv},
  primaryclass  = {hep-ph}
}

@article{Harland_Lang_2021,
	doi = {10.21468/scipostphys.11.3.064},
	year = 2021,
	volume = {11},
	number = {3},
	author = {Lucian Harland-Lang and Valeri Khoze and Mikhail G. Ryskin},
	title = {Elastic photon-initiated production at the {LHC}: the role of  hadron-hadron interactions},
	journal = {{SciPost} Phys.},
   eprint        = {2104.13392},
  archiveprefix = {arXiv},
  primaryclass  = {hep-ph}

}

@article{Sjostrand:2014zea,
      author         = "Sj{\"o}strand, Torbj{\"o}rn and Ask, Stefan and Christiansen,
                        Jesper R. and Corke, Richard and Desai, Nishita and Ilten,
                        Philip and Mrenna, Stephen and Prestel, Stefan and
                        Rasmussen, Christine O. and Skands, Peter Z.",
      title          = "{An introduction to PYTHIA 8.2}",
      journal        = "Comput. Phys. Commun.",
      volume         = "191",
      year           = "2015",
      pages          = "159",
      doi            = "10.1016/j.cpc.2015.01.024",
      eprint         = "1410.3012",
      archivePrefix  = "arXiv",
      primaryClass   = "hep-ph",
      reportNumber   = "LU-TP-14-36, MCNET-14-22, CERN-PH-TH-2014-190,
                        FERMILAB-PUB-14-316-CD, DESY-14-178, SLAC-PUB-16122,
                        --FERMILAB-PUB-14-316-CD",
      SLACcitation   = "%%CITATION = ARXIV:1410.3012;%%"
}

@Article{Cowan:2010js,
   Author = {Cowan, Glen and Cranmer, Kyle and Gross, Eilam and Vitells, Ofer},
   Title = {{Asymptotic formulae for likelihood-based tests of new physics}},
   Journal = {Eur. Phys. J. C},
   Volume = {71},
   Year = {2011},
   Pages = {1554},
   doi = "10.1140/epjc/s10052-011-1554-0",
   Eprint = "1007.1727",
   Archiveprefix = {arXiv},
   Primaryclass = {physics.data-an},
   related = "Cowan:2010js-err",
   relatedstring  = "Erratum:",
}

@Article{PERF-2007-01,
    author         = "{ATLAS Collaboration}",
    title          = "{The ATLAS Experiment at the CERN Large Hadron Collider}",
    journal        = "JINST",
    volume         = "3",
    year           = "2008",
    pages          = "S08003",
    doi            = "10.1088/1748-0221/3/08/S08003",
    primaryClass   = "hep-ex",
}

@Article{SOFT-2010-01,
    author         = "{ATLAS Collaboration}",
    title          = "{The ATLAS Simulation Infrastructure}",
    journal        = "Eur. Phys. J. C",
    volume         = "70",
    year           = "2010",
    pages          = "823",
    doi            = "10.1140/epjc/s10052-010-1429-9",
    eprint         = "1005.4568",
    archivePrefix  = "arXiv",
    primaryClass   = "physics.ins-det",
}

@Article{HIGG-2013-08,
    author         = "{ATLAS Collaboration}",
    title          = "{Measurement of Higgs boson production in the diphoton decay channel in \(pp\) collisions at center-of-mass energies of \(7\) and \(8\,\text{TeV}\) with the ATLAS detector}",
    journal        = "Phys. Rev. D",
    volume         = "90",
    year           = "2014",
    pages          = "112015",
    doi            = "10.1103/PhysRevD.90.112015",
    reportNumber   = "CERN-PH-EP-2014-198",
    eprint         = "1408.7084",
    archivePrefix  = "arXiv",
    primaryClass   = "hep-ex",
}

@Article{HIGG-2014-04,
    author         = "{ATLAS Collaboration}",
    title          = "{Search for Scalar Diphoton Resonances in the Mass Range 65--600\,GeV with the ATLAS Detector in \(pp\) Collision Data at \(\sqrt{s} = 8\,\text{TeV}\)}",
    journal        = "Phys. Rev. Lett.",
    volume         = "113",
    year           = "2014",
    pages          = "171801",
    doi            = "10.1103/PhysRevLett.113.171801",
    reportNumber   = "CERN-PH-EP-2014-142",
    eprint         = "1407.6583",
    archivePrefix  = "arXiv",
    primaryClass   = "hep-ex",
}

@Article{PERF-2014-07,
    author         = "{ATLAS Collaboration}",
    title          = "{Topological cell clustering in the ATLAS calorimeters and its performance in LHC Run~1}",
    journal        = "Eur. Phys. J. C",
    volume         = "77",
    year           = "2017",
    pages          = "490",
    doi            = "10.1140/epjc/s10052-017-5004-5",
    reportNumber   = "CERN-PH-EP-2015-304",
    eprint         = "1603.02934",
    archivePrefix  = "arXiv",
    primaryClass   = "hep-ex",
}

@Article{HIGG-2016-08,
    author         = "{ATLAS Collaboration}",
    title          = "{Search for resonances in diphoton events at \(\sqrt{s} = 13\,\text{TeV}\) with the ATLAS detector}",
    journal        = "JHEP",
    volume         = "09",
    year           = "2016",
    pages          = "001",
    doi            = "10.1007/JHEP09(2016)001",
    reportNumber   = "CERN-EP-2016-120",
    eprint         = "1606.03833",
    archivePrefix  = "arXiv",
    primaryClass   = "hep-ex",
}

@Article{HION-2016-05,
    author         = "{ATLAS Collaboration}",
    title          = "{Evidence for light-by-light scattering in heavy-ion collisions with the ATLAS detector at the LHC}",
    journal        = "Nature Phys.",
    volume         = "13",
    year           = "2017",
    pages          = "852",
    doi            = "10.1038/nphys4208",
    reportNumber   = "CERN-EP-2016-316",
    eprint         = "1702.01625",
    archivePrefix  = "arXiv",
    primaryClass   = "hep-ex",
}

@Article{PERF-2017-01,
    author         = "{ATLAS Collaboration}",
    title          = "{Electron reconstruction and identification in the ATLAS experiment using the 2015 and 2016 LHC proton--proton collision data at \(\sqrt{s} = 13\,\text{TeV}\)}",
    journal        = "Eur. Phys. J. C",
    volume         = "79",
    year           = "2019",
    pages          = "639",
    doi            = "10.1140/epjc/s10052-019-7140-6",
    reportNumber   = "CERN-EP-2018-273",
    eprint         = "1902.04655",
    archivePrefix  = "arXiv",
    primaryClass   = "hep-ex",
}

@Article{DAPR-2018-01,
    author         = "{ATLAS Collaboration}",
    title          = "{ATLAS data quality operations and performance for 2015--2018 data-taking}",
    journal        = "JINST",
    volume         = "15",
    year           = "2020",
    pages          = "P04003",
    doi            = "10.1088/1748-0221/15/04/P04003",
    reportNumber   = "CERN-EP-2019-207",
    eprint         = "1911.04632",
    archivePrefix  = "arXiv",
    primaryClass   = "physics.ins-det",
}

@Article{EGAM-2018-01,
    author         = "{ATLAS Collaboration}",
    title          = "{Electron and photon performance measurements with the ATLAS detector using the 2015--2017 LHC proton--proton collision data}",
    journal        = "JINST",
    volume         = "14",
    year           = "2019",
    pages          = "P12006",
    doi            = "10.1088/1748-0221/14/12/P12006",
    reportNumber   = "CERN-EP-2019-145",
    eprint         = "1908.00005",
    archivePrefix  = "arXiv",
    primaryClass   = "hep-ex",
}

@Article{HIGG-2018-27,
    author         = "{ATLAS Collaboration}",
    title          = "{Search for resonances decaying into photon pairs in \(139\,\text{fb}^{-1}\) of \(pp\) collisions at \(\sqrt{s} = 13\,\text{TeV}\) with the ATLAS detector}",
    journal        = "Phys. Lett. B",
    volume         = "822",
    year           = "2021",
    pages          = "136651",
    doi            = "10.1016/j.physletb.2021.136651",
    reportNumber   = "CERN-EP-2020-248",
    eprint         = "2102.13405",
    archivePrefix  = "arXiv",
    primaryClass   = "hep-ex",
}

@Article{HION-2018-19,
    author         = "{ATLAS Collaboration}",
    title          = "{Observation of Light-by-Light Scattering in Ultraperipheral Pb+Pb Collisions with the ATLAS Detector}",
    journal        = "Phys. Rev. Lett.",
    volume         = "123",
    year           = "2019",
    pages          = "052001",
    doi            = "10.1103/PhysRevLett.123.052001",
    reportNumber   = "CERN-EP-2019-051",
    eprint         = "1904.03536",
    archivePrefix  = "arXiv",
    primaryClass   = "hep-ex",
}

@Article{STDM-2018-16,
    author         = "{ATLAS Collaboration}",
    title          = "{Observation and Measurement of Forward Proton Scattering in Association with Lepton Pairs Produced via the Photon Fusion Mechanism at ATLAS}",
    journal        = "Phys. Rev. Lett.",
    volume         = "125",
    year           = "2020",
    pages          = "261801",
    doi            = "10.1103/PhysRevLett.125.261801",
    reportNumber   = "CERN-EP-2020-167",
    eprint         = "2009.14537",
    archivePrefix  = "arXiv",
    primaryClass   = "hep-ex",
}

@Article{TRIG-2018-05,
    author         = "{ATLAS Collaboration}",
    title          = "{Performance of electron and photon triggers in ATLAS during LHC Run~2}",
    journal        = "Eur. Phys. J. C",
    volume         = "80",
    year           = "2020",
    pages          = "47",
    doi            = "10.1140/epjc/s10052-019-7500-2",
    reportNumber   = "CERN-EP-2019-169",
    eprint         = "1909.00761",
    archivePrefix  = "arXiv",
    primaryClass   = "hep-ex",
}

@Article{HION-2019-08,
    author         = "{ATLAS Collaboration}",
    title          = "{Measurement of light-by-light scattering and search for axion-like particles with \(2.2\,\text{nb}^{-1}\) of Pb+Pb data with the ATLAS detector}",
    journal        = "JHEP",
    volume         = "03",
    year           = "2021",
    pages          = "243",
    doi            = "10.1007/JHEP03(2021)243",
    reportNumber   = "CERN-EP-2020-135",
    eprint         = "2008.05355",
    archivePrefix  = "arXiv",
    primaryClass   = "hep-ex",
    related        = "HION-2019-08-err",
    relatedstring  = "Erratum:",
}

@Booklet{ATL-SOFT-PUB-2021-001,
    author         = "{ATLAS Collaboration}",
    title          = "{The ATLAS Collaboration Software and Firmware}",
    howpublished   = "{ATL-SOFT-PUB-2021-001}",
    url            = "https://cds.cern.ch/record/2767187",
    year           = "2021",
}

@Booklet{ATL-SOFT-PUB-2021-003,
    author         = "{ATLAS Collaboration}",
    title          = "{ATLAS Computing Acknowledgements}",
    howpublished   = "{ATL-SOFT-PUB-2021-003}",
    url            = "https://cds.cern.ch/record/2776662",
    year           = "2021",
}

@Report{ATLAS-TDR-18,
    author         = "{ATLAS Collaboration}",
    title          = "{ATLAS Forward Detectors for Measurement of Elastic Scattering and Luminosity: Technical Design Report}",
    type           = "ATLAS-TDR-18; CERN-LHCC-2008-004",
    year           = "2008",
    url            = "https://cds.cern.ch/record/1095847",
}

@Report{ATLAS-TDR-24,
    author         = "{ATLAS Collaboration}",
    title          = "{ATLAS Forward Proton Phase-I Upgrade: Technical Design Report}",
    type           = "ATLAS-TDR-024; CERN-LHCC-2015-009",
    year           = "2015",
    url            = "https://cds.cern.ch/record/2017378",
}

@Article{CMS-FSQ-16-012,
    author         = "{CMS Collaboration}",
    title          = "{Evidence for light-by-light scattering and searches for axion-like particles in ultraperipheral PbPb collisions at \(\sqrt{s_{\text{NN}}} = 5.02\,\text{TeV}\)}",
    journal        = "Phys. Lett. B",
    volume         = "797",
    year           = "2019",
    pages          = "134826",
    doi            = "10.1016/j.physletb.2019.134826",
    reportNumber   = "CERN-EP-2018-271",
    eprint         = "1810.04602",
    archivePrefix  = "arXiv",
    primaryClass   = "hep-ex",
}

@Article{CMS-PPS-17-001,
    author         = "{CMS Collaboration}",
    title          = "{Observation of proton-tagged, central (semi)exclusive production of high-mass lepton pairs in \(pp\) collisions at \(13\,\text{TeV}\) with the CMS-TOTEM precision proton spectrometer}",
    journal        = "JHEP",
    volume         = "07",
    year           = "2018",
    pages          = "153",
    doi            = "10.1007/JHEP07(2018)153",
    reportNumber   = "CERN-EP-2018-014",
    eprint         = "1803.04496",
    archivePrefix  = "arXiv",
    primaryClass   = "hep-ex",
}

@Article{CMS-EXO-18-014,
    author         = "{CMS Collaboration}",
    title          = "{First Search for Exclusive Diphoton Production at High Mass with Tagged Protons in Proton--Proton Collisions at \(\sqrt{s} = 13\,\text{TeV}\)}",
    journal        = "Phys. Rev. Lett.",
    volume         = "129",
    year           = "2021",
    pages          = "011801",
    doi            = "10.1103/PhysRevLett.129.011801",
    reportNumber   = "CERN-EP-2021-191",
    eprint         = "2110.05916",
    archivePrefix  = "arXiv",
    primaryClass   = "hep-ex",
}

@Booklet{ATL-PHYS-PUB-2010-013,
    author         = "{ATLAS Collaboration}",
    title          = "{The simulation principle and performance of the ATLAS fast calorimeter simulation FastCaloSim}",
    howpublished   = "{ATL-PHYS-PUB-2010-013}",
    url            = "https://cds.cern.ch/record/1300517",
    year           = "2010",
}

@Booklet{ATL-PHYS-PUB-2016-017,
    author         = "{ATLAS Collaboration}",
    title          = "{The Pythia~8 A3 tune description of ATLAS minimum bias and inelastic measurements incorporating the Donnachie--Landshoff diffractive model}",
    howpublished   = "{ATL-PHYS-PUB-2016-017}",
    url            = "https://cds.cern.ch/record/2206965",
    year           = "2016",
}

@Booklet{ATL-PHYS-PUB-2017-012,
    author         = "{ATLAS Collaboration}",
    title          = "{Proton tagging with the one arm AFP detector}",
    howpublished   = "{ATL-PHYS-PUB-2017-012}",
    url            = "https://cds.cern.ch/record/2273274",
    year           = "2017",
}

@Booklet{ATL-DAQ-PUB-2018-002,
    author         = "{ATLAS Collaboration}",
    title          = "{Trigger Menu in 2017}",
    howpublished   = "{ATL-DAQ-PUB-2018-002}",
    url            = "https://cds.cern.ch/record/2625986",
    year           = "2018",
}
 
\clearpage
% ATLAS Collaboration author list
% Reference date of EXOT-2019-28 is 2023-02-08
% Author list last updated on date 24-JUN-23
% Data extracted on 24-Jun-2023 for paper reference EXOT-2019-28
% at 11:46am
 
\begin{flushleft}
\hypersetup{urlcolor=black}
{\Large The ATLAS Collaboration}

\bigskip

\AtlasOrcid[0000-0002-6665-4934]{G.~Aad}$^\textrm{\scriptsize 102}$,
\AtlasOrcid[0000-0002-5888-2734]{B.~Abbott}$^\textrm{\scriptsize 120}$,
\AtlasOrcid[0000-0002-1002-1652]{K.~Abeling}$^\textrm{\scriptsize 55}$,
\AtlasOrcid[0000-0001-5763-2760]{N.J.~Abicht}$^\textrm{\scriptsize 49}$,
\AtlasOrcid[0000-0002-8496-9294]{S.H.~Abidi}$^\textrm{\scriptsize 29}$,
\AtlasOrcid[0000-0002-9987-2292]{A.~Aboulhorma}$^\textrm{\scriptsize 35e}$,
\AtlasOrcid[0000-0001-5329-6640]{H.~Abramowicz}$^\textrm{\scriptsize 151}$,
\AtlasOrcid[0000-0002-1599-2896]{H.~Abreu}$^\textrm{\scriptsize 150}$,
\AtlasOrcid[0000-0003-0403-3697]{Y.~Abulaiti}$^\textrm{\scriptsize 117}$,
\AtlasOrcid[0000-0003-0762-7204]{A.C.~Abusleme~Hoffman}$^\textrm{\scriptsize 137a}$
% ...
\bigskip
\\

% ...
$^{29}$Physics Department, Brookhaven National Laboratory, Upton NY; United States of America.\\
% ...
$^{35}$$^{(a)}$Facult\'e des Sciences Ain Chock, R\'eseau Universitaire de Physique des Hautes Energies - Universit\'e Hassan II, Casablanca;$^{(b)}$Facult\'{e} des Sciences, Universit\'{e} Ibn-Tofail, K\'{e}nitra;$^{(c)}$Facult\'e des Sciences Semlalia, Universit\'e Cadi Ayyad, LPHEA-Marrakech;$^{(d)}$LPMR, Facult\'e des Sciences, Universit\'e Mohamed Premier, Oujda;$^{(e)}$Facult\'e des sciences, Universit\'e Mohammed V, Rabat;$^{(f)}$Institute of Applied Physics, Mohammed VI Polytechnic University, Ben Guerir; Morocco.\\
% ...
$^{49}$Fakult\"{a}t Physik , Technische Universit{\"a}t Dortmund, Dortmund; Germany.\\
% ...
$^{55}$II. Physikalisches Institut, Georg-August-Universit\"{a}t G\"ottingen, G\"ottingen; Germany.\\
% ...
$^{102}$CPPM, Aix-Marseille Universit\'e, CNRS/IN2P3, Marseille; France.\\
% ...
$^{117}$Department of Physics, New York University, New York NY; United States of America.\\
% ...
$^{120}$Homer L. Dodge Department of Physics and Astronomy, University of Oklahoma, Norman OK; United States of America.\\
% ...
$^{137}$$^{(a)}$Departamento de F\'isica, Pontificia Universidad Cat\'olica de Chile, Santiago;$^{(b)}$Millennium Institute for Subatomic physics at high energy frontier (SAPHIR), Santiago;$^{(c)}$Instituto de Investigaci\'on Multidisciplinario en Ciencia y Tecnolog\'ia, y Departamento de F\'isica, Universidad de La Serena;$^{(d)}$Universidad Andres Bello, Department of Physics, Santiago;$^{(e)}$Instituto de Alta Investigaci\'on, Universidad de Tarapac\'a, Arica;$^{(f)}$Departamento de F\'isica, Universidad T\'ecnica Federico Santa Mar\'ia, Valpara\'iso; Chile.\\
% ...
$^{150}$Department of Physics, Technion, Israel Institute of Technology, Haifa; Israel.\\
$^{151}$Raymond and Beverly Sackler School of Physics and Astronomy, Tel Aviv University, Tel Aviv; Israel.\\
% ...
$^{a}$ Also Affiliated with an institute covered by a cooperation agreement with CERN.\\
$^{b}$ Also at An-Najah National University, Nablus; Palestine.\\
$^{c}$ Also at APC, Universit\'e Paris Cit\'e, CNRS/IN2P3, Paris; France.\\
$^{d}$ Also at Borough of Manhattan Community College, City University of New York, New York NY; United States of America.\\
$^{e}$ Also at Center for High Energy Physics, Peking University; China.\\
$^{f}$ Also at Center for Interdisciplinary Research and Innovation (CIRI-AUTH), Thessaloniki ; Greece.\\
% ...

\end{flushleft}

% Created with Glance <Atlas.Glance@cern.ch>

\FloatBarrier

\end{document}